\PassOptionsToPackage{pdfpagelabels=false}{hyperref}
\documentclass[fleqn,usenatbib,usedcolumn]{mnras}
\usepackage[british]{babel}             % British English hyphenation
\usepackage{txfonts}                  % Good fonts

\usepackage{graphicx}	% Including figure files
\usepackage{hyperref}	% hyperlinks
\usepackage{soul}
\usepackage{bm}

\newcommand*\rad{~rad\,m$^{-2}$}

\DeclareRobustCommand{\ion}[2]{%
\relax\ifmmode
\ifx\testbx\f@series
{\mathbf{#1\,\mathsc{#2}}}\else
{\mathrm{#1\,\mathsc{#2}}}\fi
\else\textup{#1\,{\mdseries\textsc{#2}}}%
\fi}

\title[LOFAR DR2 RM Grid]{The Faraday Rotation Measure Grid of the LOFAR Two-metre Sky Survey: Data Release 2}
\author[O'Sullivan et al.]{\parbox{\textwidth}{
S.~P.~O'Sullivan,$^{1}$\thanks{shane.osullivan@dcu.ie}
T.~W.~Shimwell$^{2,3}$, M.~J.~Hardcastle$^{4}$, C.~Tasse$^{5,6}$, G.~Heald$^{7}$, E.~Carretti$^{8}$, M.~Br\"uggen$^{9}$, V.~Vacca$^{10}$, C.~Sobey$^{7}$, C.~L.~Van Eck$^{11}$, C.~Horellou$^{12}$, R.~Beck$^{13}$, M.~Bilicki$^{14}$, S.~Bourke$^{12,15}$,  A.~Botteon$^{3}$,  J.~H.~Croston$^{16}$, A.~Drabent$^{17}$, K.~Duncan$^{18}$,  V.~Heesen$^{9}$, S.~Ideguchi$^{19}$, M.~Kirwan$^{1}$, L.~Lawlor$^{1}$, B.~Mingo$^{16}$, B.~Nikiel-Wroczy\'nski$^{20}$, J.~Piotrowska$^{20}$, A.~M.~M.~Scaife$^{21,22}$, R.~J.~van Weeren$^{3}$ \newline
\emph{\normalsize Affiliations are listed at the end of the paper}
}}

\begin{document}

\date{Accepted 2022 December 22. Received 2022 December 19; in original form 2022 May 30.}

\pagerange{\pageref{firstpage}--\pageref{lastpage}} \pubyear{2023}

\maketitle

\label{firstpage}

\begin{abstract}
%background, aims, methods, results, discussion, conclusions 
A Faraday rotation measure (RM) catalogue, or RM Grid, is a valuable resource for the study of cosmic magnetism. 
Using the second data release (DR2) from the LOFAR Two-metre Sky Survey (LoTSS), we have produced a catalogue of 2461 extragalactic high-precision 
RM values across 5720~deg$^{2}$ of sky (corresponding to a polarized source areal number density of $\sim$0.43~deg$^{-2}$). 
The linear polarization and RM properties were derived using RM synthesis from the 
Stokes $Q$ and $U$ channel images at an angular resolution of 20\arcsec~across a frequency range of 120 to 168~MHz 
with a channel bandwidth of 97.6 kHz. The fraction of total intensity sources ($>1$~mJy~beam$^{-1}$) found to be polarized was $\sim$0.2\%. 
The median detection threshold was 0.6~mJy~beam$^{-1}$ ($8\sigma_{QU}$), with 
a median RM uncertainty of 0.06\rad~{(although a systematic uncertainty of up to 0.3\rad~is possible, after the ionosphere RM correction)}. 
The median degree of polarization of the detected sources is 1.8\%, with a range of 0.05\% to 31\%. 
Comparisons with cm-wavelength RMs indicate minimal amounts of Faraday complexity in the LoTSS detections, making them 
ideal sources for RM Grid studies. 
Host galaxy identifications were obtained for 88\% of the sources, along with redshifts for 
79\% (both photometric and spectroscopic), with the median redshift being 0.6. 
The focus of the current catalogue was on reliability rather than completeness, 
and we expect future versions of the LoTSS RM Grid to have a higher areal number density. 
In addition, 25 pulsars were identified, mainly through their high degrees of linear polarization. 

\end{abstract}

\begin{keywords}
techniques: polarimetric -- galaxies:active 
\end{keywords}

\section{Introduction}\label{sec:intro}

The construction of large-area `RM Grid' catalogues are a key goal for current and future radio telescopes \citep[e.g.][]{gaenslerSKA2004,possum,lacy2020}. The RM Grid is shorthand for a collection of Faraday rotation measure (RM) values from linearly polarized radio sources observed across a particular area of sky, and it enables many different science goals in the study of magnetic fields on different scales in the Universe \citep{beckgaensler2004,johnstonhollitt2015,heald2020}. 

While most RM Grid studies to date have been at centimetre wavelengths \citep[e.g.][]{taylor2009}, the recent development of radio facilities at low frequencies has led to a significant advance in our understanding of the population of polarized radio sources and their Faraday rotation properties \citep{Bernardi:2013, Mulcahy:2014, Jelic:2015, orru2015, Lenc:2016,riseley2018,osullivanlenc2018,vaneck2018,neld2018,riseley2020}, and how they can be used to enhance our understanding of magnetic fields in different cosmic environments \citep[e.g.][]{osullivan2019,osullivan2020,cantwell2020,stuardi2020,mahatma2021,carretti2022}. 

The main advantage of RM studies at metre wavelengths compared to centimetre wavelength observations is the dramatic improvement in the accuracy with which individual RM values can be determined (by one to two orders of magnitude). However, one of the main challenges to finding linearly polarized sources at long wavelengths is the need for high angular resolution and high sensitivity observations to mitigate the strong influence of Faraday depolarization. These challenges are being met to a large degree by LOFAR with its unique ability to produce high fidelity images at high angular resolution, in principle as high as 0.3\arcsec~\citep{morabito2016,jackson2016,harris2019,sweijen2022}. Furthermore, the wide field of view and frequency bandwidth enables large areas of sky to be covered efficiently in order to detect many linearly polarized sources and their associated RM values. 

The combination of RM Grid catalogues at metre and centimetre wavelengths provide an important means to better understand the different contributions to the Faraday rotation along the line of sight. For example, radio source populations that are located in, or have lines of sight through, dense magnetoionic environments are strongly affected by Faraday depolarization at long wavelengths \citep[e.g., LOFAR;][]{vanhaarlem2013} 
%\citep[LOFAR, MWA:][]{vanhaarlem2013,Tingay:2013} 
but less so at shorter wavelengths \citep[e.g., ASKAP-POSSUM;][]{possum}. 
%\citep[e.g., ASKAP-POSSUM, VLASS:][]{possum,lacy2020}. 
These observations thus provide additional constraints for models that attempt to isolate the RM contribution in the radio source's local environment from that due to the intergalactic medium and the Milky Way, for example. 

Classifications of the source properties such as the host galaxy, redshift, morphology, etc.~are important to identify and study the different underlying populations, as well as providing a means of better statistical weighting to optimise the inferences for particular science goals \citep{rudnick2019, vacca2016}. In addition to the RM Grid catalogue, the ongoing LOFAR Two-metre Sky Survey \citep[LoTSS;][]{shimwell2017,shimwell2019,shimwell2022} provides a wealth of added-value data products, with host galaxy identifications \citep{williams2019}, photometric redshift estimates \citep{duncan2019}, spectroscopic redshift observations \citep{smith2016}, source morphology and environment classifications of radio galaxies \citep{mingo2019, hardcastle2019, croston2019}, and star forming galaxy scaling relations \citep{smith2021,heesen2022}.  

In this paper, we use the LoTSS polarization data at an angular resolution of 20\arcsec. LoTSS is observing the northern sky with the LOFAR High-Band Antennas (HBA) at Declinations greater than 0\degr~with a frequency range of 120 to 168~MHz.  
As part of Data Release 2 \citep[DR2;][]{shimwell2022}, here we present the LoTSS-DR2 RM Grid covering 5,720~deg$^{2}$~of the sky. This is approximately a quarter of the final sky area expected from the full LoTSS survey. In addition to the RM Grid catalogue, we also provide access to a wide range of ancillary data products, such as RM cubes and Stokes $Q$ and $U$ frequency spectra (see Data Availability section, prior to the References). In our description of the catalogue construction, we also highlight some of the limitations of the current data products for science. As we are continually developing our calibration and data analysis tools for long wavelength polarimetry, we expect that future data releases will provide significant improvements in the number of detected polarized sources and also in the overall data quality which will allow for more detailed scientific studies of individual sources. 

In Section~\ref{sec:data}, we describe the observational data and our polarized source detection algorithm. 
The RM Grid catalogue is presented in Section~\ref{sec:results}, along with the description of the value-added products. 
In Section~\ref{sec:summary}, we provide a summary of the main results and a perspective on enhancements planned for future data releases. 
%Throughout this paper, we assume a $\Lambda$CDM cosmology with 
%H$_0 = 67.8$ km s$^{-1}$ Mpc$^{-1}$, $\Omega_M=0.308$ and $\Omega_{\Lambda}=0.692$ \citep{planck2016xiii}.

\subsection{Linear polarization and Faraday rotation definitions}

The complex linear polarization is defined as 
\begin{equation} \label{complexP}
P = Q+iU = p I e^{2i \psi}
\end{equation}

\noindent where $I$, $Q$, $U$ are the Stokes parameters, $p$ is the degree 
of polarization and $\psi$ is the observed polarization angle. The angle $\psi$ 
rotates linearly with wavelength-squared from its intrinsic value $\psi_0$ as it 
propagates through foreground regions of magnetised plasma due to the effect 
of Faraday rotation, i.e.~
\begin{equation} \label{RM}
\psi=\psi_0+\phi\,\lambda^2,
\end{equation}
where $\phi$ is the Faraday depth, defined as

\begin{equation} \label{FaradayDepth}
%\phi(L) = 0.812 \int^{\rm observer}_{\rm source}{ n_e \bmath{B}\cdot d\bmath{l}} ~~{\rm rad~m}^{-2}.
\phi(L) = 0.812 \int^{L}_{0}{ n_e\, {B_{||}} \,dl} \,\,\,~~{\rm rad~m}^{-2}.
\end{equation}

\noindent Here $n_e$ is the free electron density (in units of cm$^{-3}$), 
$B_{||}$ is the line-of-sight magnetic field strength (in $\mu$G) and $dl$ is the infinitesimal path length (in parsecs). 

Due to the finite angular resolution of our telescopes, the observed complex linear 
polarization intensity, $\bm{P}(\lambda^2)$, is effectively the sum of the polarized emission 
from all Faraday depths within the synthesised beam, with  
\begin{equation}
\bm{P}(\lambda^2)=\int_{-\infty}^{\infty} \bm{F}(\phi) \, e^{2i\phi\lambda^2} \, d\phi,
\end{equation}
where $\bm{F}(\phi)$ describes the distribution of polarized emission as a function 
of Faraday depth. $\bm{F}(\phi)$ is called the Faraday dispersion function (FDF) or Faraday depth spectrum.  

In order to identify linearly polarized radio sources, we employ the technique of RM synthesis \citep{burn1966,bdb2005} 
where one takes the Fourier transform of $\bm{P}(\lambda^2)$ to estimate the FDF. 
In the simplest case, where there is a single background source of polarized 
emission encountering Faraday rotation in the foreground, 
the Faraday depth and the Faraday rotation measure (RM) are equivalent. 
In this work, we determine the RM of the sources from the Faraday depth of the peak of $|\bm{F}(\phi)|$. 

\section{Data analysis}\label{sec:data}

The LoTSS-DR2 sky area imaged in polarization covers 5720~deg$^{2}$ which was split between two fields: the 13~hr field, a contiguous area of 4240~deg$^{2}$ centred at a Right Ascension (RA) of approximately 13$^{\rm h}$00$^{\rm m}$ and a Declination (Dec.) of $+47^\circ$, and the 0~hr field of 1480~deg$^{2}$ centred at an RA of approximately 00$^{\rm h}$30$^{\rm m}$ and a Dec.~of $+30^\circ$. The 13~hr field is composed of 626 pointings (i.e.~8~hr integrations) and the 0~hr field has 215 pointings. 

\subsection{Initial LoTSS data products}
After the initial direction-independent calibration steps using {\sc prefactor}\footnote{https://github.com/lofar-astron/prefactor}, the LoTSS data undergo a facet-based direction-dependent calibration using {\sc killms} and {\sc ddfacet} \citep[see][for the details]{shimwell2019,tasse2021}, run by the DR2 {\sc ddf-pipeline}\footnote{https://github.com/mhardcastle/ddf-pipeline}. This pipeline outputs Stokes $I$ images at 6\arcsec~and 20\arcsec~resolution, as well as Stokes $Q$ and $U$ image cubes at 20\arcsec~and 3\arcmin~(with 480 images per Stokes parameter, a channel image bandwidth of 97.6~kHz, across a frequency range from $\sim$120 to 168 MHz using the HBA). 
In this work, we use the 20\arcsec~data in addition to the Stokes $I$ source catalogues and images produced by the LoTSS team \citep[e.g.][]{williams2019,shimwell2022}. The 3\arcmin~$QU$ cubes and Stokes $V$ images are used elsewhere \citep[e.g.][]{vaneck2019, erceg2022, callingham2020}.  

A 20\arcsec~$QU$ image cube for each LoTSS field covers an area of $\sim$58~deg$^{2}$~(7.6\degr~x 7.6\degr). The $Q$ and $U$ images are not deconvolved (due to computational and software limitations) and the cubes are compressed using the {\sc fpack} software\footnote{https://heasarc.gsfc.nasa.gov/fitsio/fpack/} by a factor of 6.4 (the default). 
The distance between LoTSS field centres is $\sim$2.58\degr~and the FWHM of the HBA station beam at 144 MHz is 3.96\degr~\citep{shimwell2017}. For our current work, we considered it unnecessary to run RM synthesis on the full field, so we extracted a smaller 4\degr~x 4\degr~area region around the centre of the $QU$ cube (after unpacking the $QU$ cube to its original state using {\sc funpack}). This choice gave us full coverage of the LoTSS sky area with no gaps between fields, while also providing significant overlap such that several polarized sources were detected in multiple adjacent fields. No mosaicking of the $QU$ cubes was attempted, as this would have introduced unnecessary depolarization due to the lack of an absolute polarization angle calibration for each field \citep[cf.][]{herreraruiz2021}. 

\begin{figure}
\includegraphics[width=8.5cm,clip=true,trim=0.2cm 0.25cm 0.0cm 0.0cm]{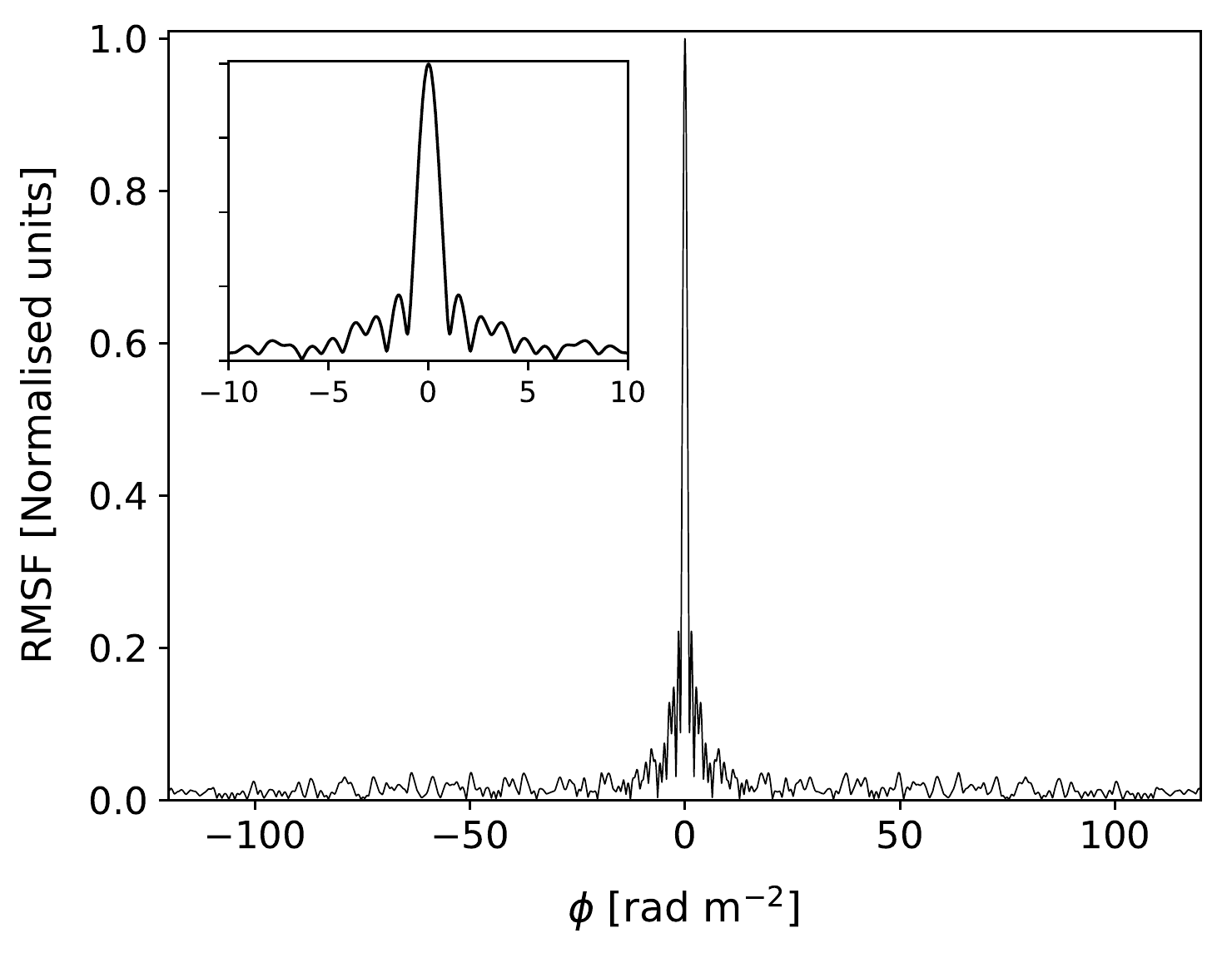}
\caption{The LoTSS-DR2 rotation measure spread function (RMSF). Obtained from an observation with 450 frequency channels (typical) and using inverse-variance weighting of the channels to produce an RMSF with a FWHM of 1.157\rad, derived from a Gaussian fit to the main lobe. The inset shows a zoom-in of $\pm$10\rad, to more clearly see the sidelobes, the first of which are $\sim$22\% of the main peak. }\label{fig:rmsf}
\end{figure}

\begin{figure}
\includegraphics[width=8.5cm,clip=true,trim=0.0cm 0.0cm 0.0cm 0.0cm]{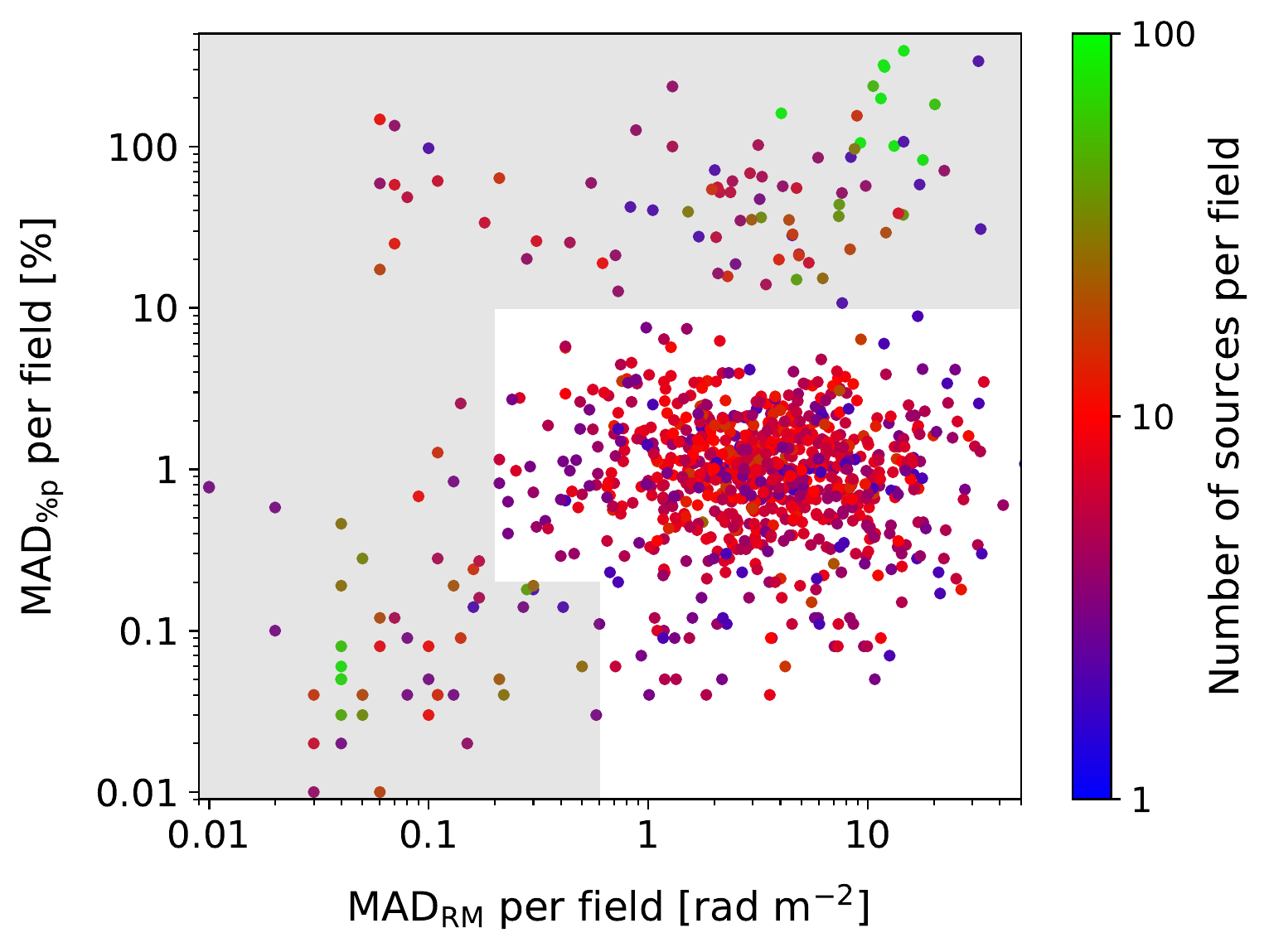}
\caption{The median absolute deviation (MAD) of the degree of polarization (MAD$_{\%p}$) and of the RM (MAD$_{\rm RM}$) for sources within each LoTSS field (i.e.~a 16~deg$^2$ region). The colour scale indicates the number of polarized sources per field above 8$\sigma_{QU}$. The fields affected by the `fake' polarized source issue are in the bottom left corner of the plot (Section~\ref{sec:fake}). The grey shaded region highlights the fields that were automatically excluded from the RM Grid catalogue (see Section~\ref{sec:cuts} for further details). }\label{fig:madRMmadfpol}
\end{figure}

\subsection{RM and polarization data products}\label{sec:original}
The LoTSS frequency range and channel bandwidth provide a typical resolution in Faraday depth space of $\sim$1.16\rad~(i.e.~the FWHM of the rotation measure spread function, RMSF, shown in Fig.~\ref{fig:rmsf}), with a maximum/minimum observable Faraday depth of $\pm$170\rad~(at full sensitivity) and  $\pm$450\rad~(at 50\% sensitivity). The largest observable scale is 0.97\rad, which is smaller than the Faraday depth resolution, implying that any observed polarized emission is unresolved in Faraday depth space. 

In order to find linearly polarized sources, 
the RM synthesis technique 
was applied on the $Q$ and $U$ images using {\sc pyrmsynth}\footnote{https://github.com/mrbell/pyrmsynth} with uniform weighting, for pixels where the 20\arcsec~total intensity was greater than 1~mJy~beam$^{-1}$. %} 
The input $Q$ and $U$ channel images were flagged if the noise in either channel image was greater than 5 times the median noise of all channels. The Faraday depth range in RM synthesis was limited to $\pm$120\rad~with a sampling of 0.3\rad, and {\sc rmclean} \citep{heald2009} was not run. These sub-optimal choices were dictated mainly by the computing and storage resources available at the time, coupled with the desire to process large areas of the sky in an efficient manner (see Section~\ref{sec:future} for future planned enhancements). 

For each pixel in the output cube of the Faraday dispersion function (FDF) or Faraday depth spectrum, we identified the peak polarized intensity outside of a user-specified instrumental polarization or `leakage' range of $-3$\rad~to $+1$\rad,\footnote{Initially the leakage exclusion range was symmetric, using $\pm1$\rad. However, after finding many more leakage peaks at negative RM values this range was extended. In future, a more robust leakage mitigation strategy that includes knowledge of the ionospheric RM correction values should help.} while also retaining a record of the highest peak in the full Faraday depth spectrum. The leakage signal (i.e.~the instrumental polarization) occurs intrinsically at 0\rad~with a degree of polarization of order 1\% of the Stokes $I$ intensity in the worst affected regions, with a typical leakage of 0.2\% \citep{shimwell2022}. However, this leakage signal is smeared out by the ionospheric RM correction, the magnitude of which is $O$(1\rad). The narrow RMSF of LoTSS ($\sim$1.16\rad) means that we can still identify a real polarized signal at low Faraday depths, just typically not within the leakage range specified above (an exception would be a source with a degree of polarization $\gg$1\%). 

To provide an initial list of detections, we estimated the noise ($\sigma_{QU}$) for each pixel in the Faraday depth cube from the rms of the wings of the real and imaginary parts of the FDF (i.e.~$>+100$\rad~and $<-100$\rad). This is likely to be a slight over-estimate of the noise for bright sources and/or sources with a large $|$RM$|$ because we did not apply {\sc rmclean}. 
For each field, we recorded the polarized intensity peaks in the FDF greater than $5.5\sigma_{QU}$. We fit a parabola to this peak in order to estimate the peak polarized intensity and the corresponding Faraday depth value (i.e.~the RM) to a higher precision than the 0.3\rad~sampling. This is used to create an RM image, a polarized intensity image, and a degree of polarization map (using the 20\arcsec~total intensity map which is on the same pixel grid). The polarized intensity image was corrected for the polarization bias following \cite{george2012}. We also output a $\sigma_{QU}$ noise image and an RM error image (calculated in the standard manner as the FWHM of the RMSF divided by twice the signal to noise ratio). We note that the total RM error budget is dominated by the residual error in 
the ionospheric RM correction, which is applied to the data with {\sc prefactor} using the {\sc rmextract} code \citep{mevius2018}. This systematic error, for an individual LoTSS pointing, is estimated to be 
$\sim$0.1 to 0.3\rad~across a typical 8~hr LoTSS observation \citep{sotomayor2013}, and is not included in the RM error image. 
In {\sc rmextract}, the time and direction-dependent ionospheric Faraday rotation is approximated by a thin shell model, using the measured total electron content (TEC) in the ionosphere and a projection of the geomagnetic field along a particular line of sight. The ionosphere TEC data are taken from  
Global Navigation Satellite System (GNSS) observations %Centre for Orbital Determination in Europe (CODE) observations 
\citep[e.g.][]{GNSS} and the geomagnetic field model is based on the World Magnetic Model (WMM) software \citep{chulliat_usuk_2020}. 

\begin{figure*}
\includegraphics[width=17.5cm,clip=true,trim=0.0cm 0.2cm 0.0cm 0.0cm]{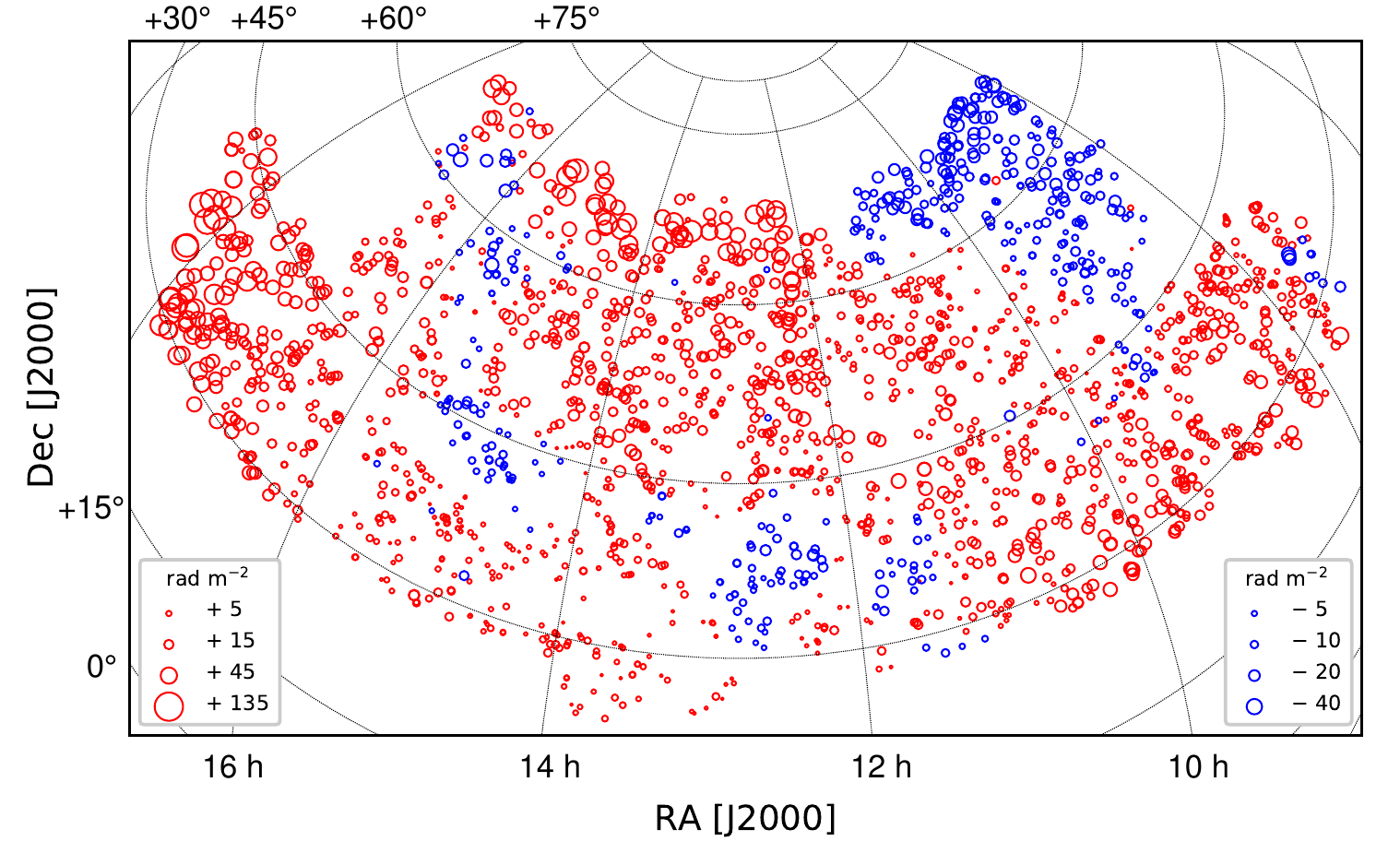}
\caption{ Sky distribution of the polarized sources across the LoTSS-DR2 13~hr field, in equatorial coordinates and with an orthographic projection. The red/blue coloured circles correspond to positive/negative RM values, and the size of the circles are proportional to the magnitude of the RM (as quantified in the figure legends). The are 2,039 sources in the 13~hr field, which corresponds to an areal number density of 0.48~deg$^{-2}$. }\label{fig:10}
\end{figure*}

\begin{figure}
\includegraphics[width=8.5cm,clip=true,trim=0.0cm 0.3cm 0.0cm 0.0cm]{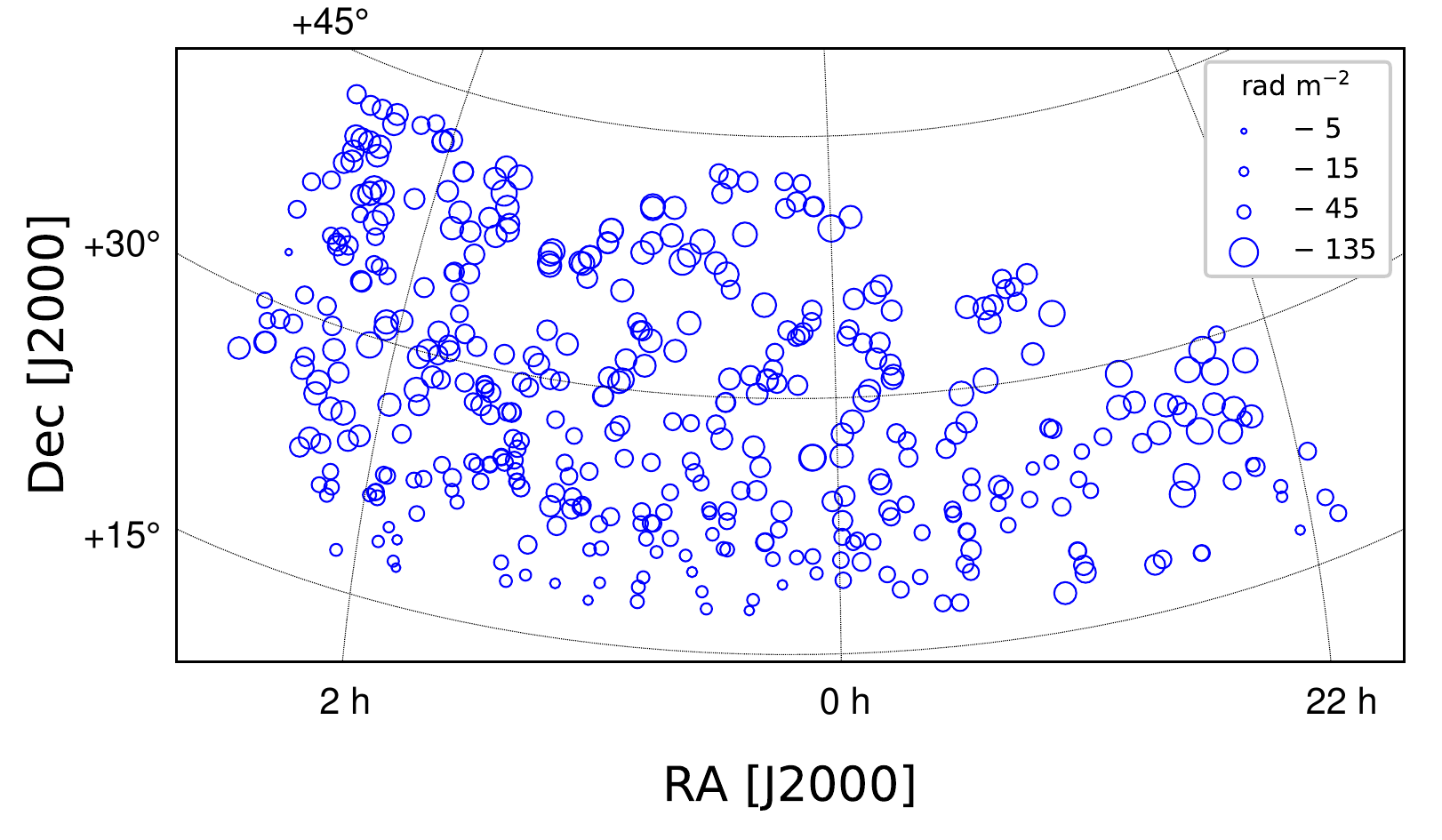}
\caption{ In a similar manner to Fig.~\ref{fig:10}, this shows the sky distribution of the polarized sources across the LoTSS-DR2 0~hr field. The are 422 sources in the 0~hr field, which corresponds to an areal number density of 0.29~deg$^{-2}$. }\label{fig:11}
\end{figure}

\begin{figure*}
\includegraphics[width=17.5cm,clip=true,trim=0.0cm 0.0cm 0.0cm 0.0cm]{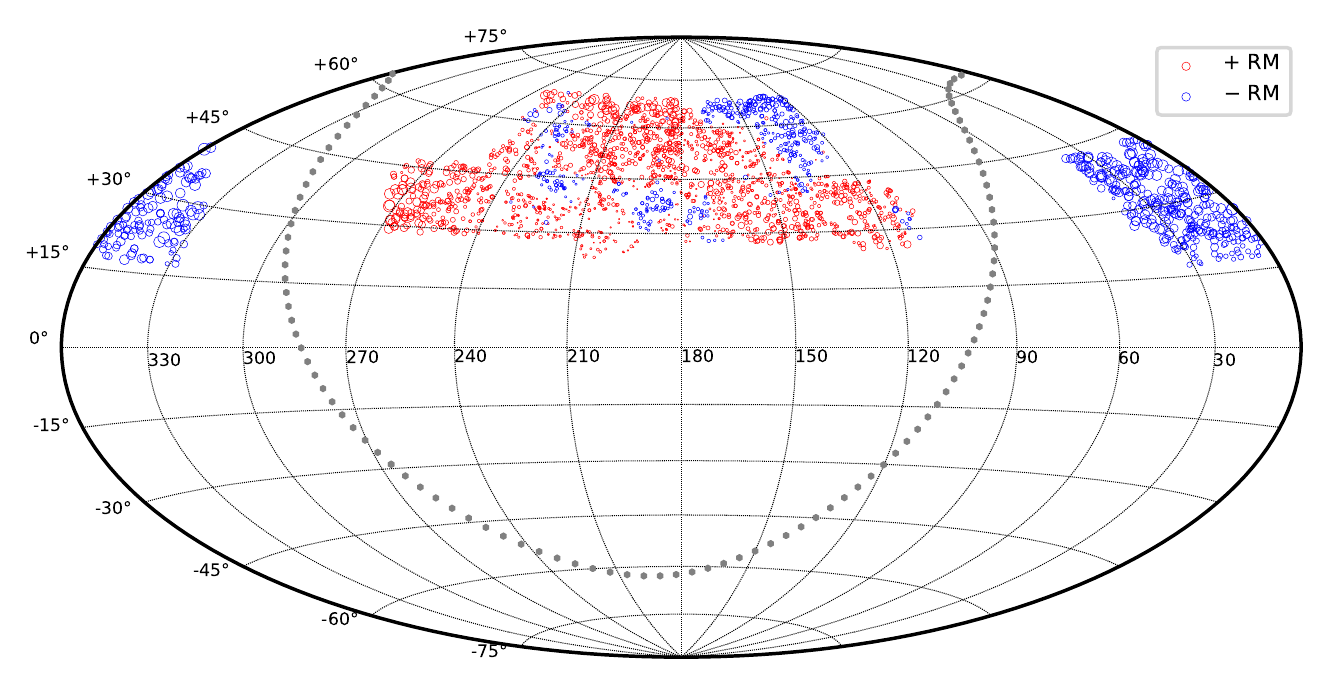}
\caption{ The LoTSS-DR2 RM Grid in equatorial coordinates with a Hammer-Aitoff equal-area projection, where the red/blue coloured circles correspond to positive/negative RM values. The 13~hr field is in the centre and the 0~hr field is split across either side. The locus of the Galactic plane is highlighted by the solid grey hexagon symbols. }\label{fig:12}
\end{figure*}

\begin{figure*}
\includegraphics[width=17.5cm,clip=true,trim=0.0cm 0.0cm 0.0cm 0.0cm]{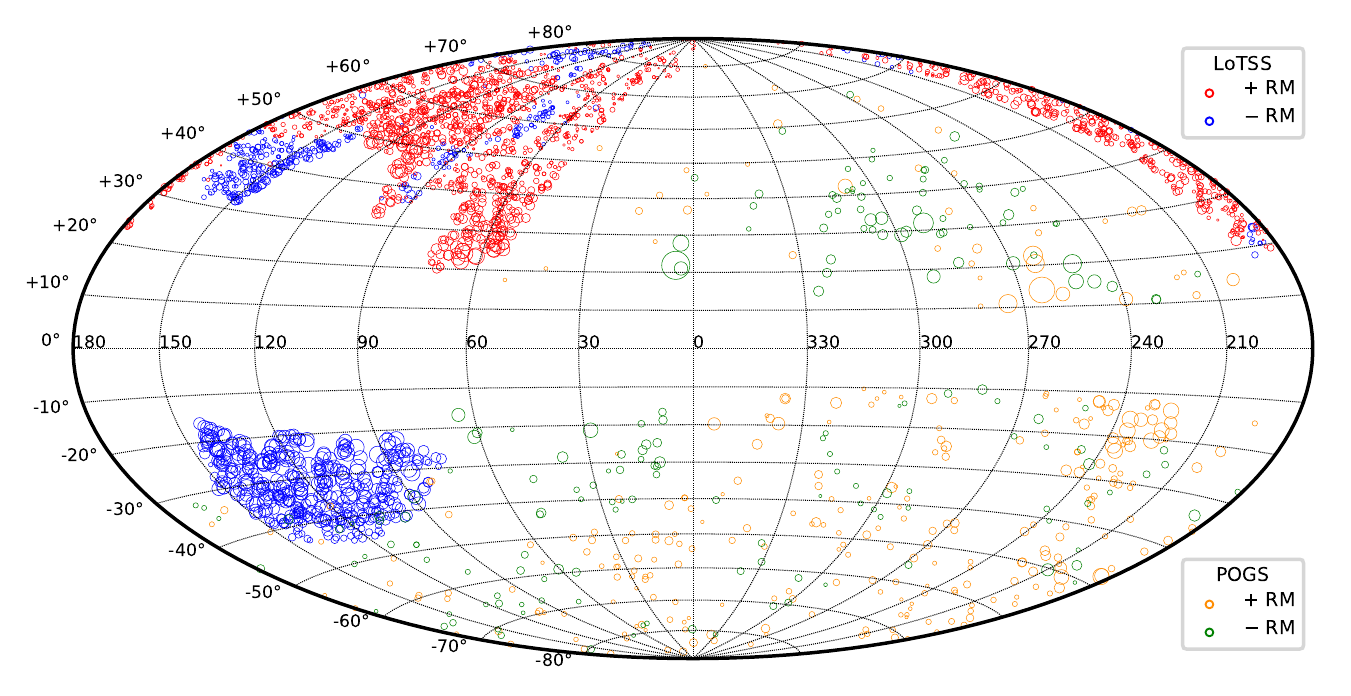}
\caption{ The LoTSS-DR2 RM Grid in Galactic coordinates with a Hammer-Aitoff equal-area projection, where the red/blue coloured circles correspond to positive/negative LoTSS RM values. The complementarity of the POGS RM Grid \citep{riseley2020} is highlighted by the difference in sky coverage, with orange/green circles corresponding to positive/negative POGS RM values.  }\label{fig:13}
\end{figure*}

\subsection{RM Grid catalogue construction}\label{sec:leakage2}

For each field, we used the polarized intensity image (created as described above) to identify candidate polarized source components. We did this using a flood-fill source finding algorithm, with a 20 pixel x 20 pixel box, where all pixels within this box with a polarized intensity $>5.5\sigma_{QU}$ were grouped together. The highest signal to noise pixel in this group was then recorded as the sky position for the catalogued values of this source component. As each pixel is 4.5\arcsec, this means that any two pixels that are separated by less than 1.5\arcmin~will be merged into a single source component. Since LoTSS polarized sources are typically sparse on the sky ($\sim$1 per 2 or 3~deg$^{2}$) this approach worked well for minimising the number of candidate polarized source components in these non-deconvolved images for further processing. However, it is worth noting that with this algorithm many double-lobed radio galaxies in which both lobes are polarized have only the brightest lobe catalogued. Close, random radio source pairs may also suffer from this, but such pairs are not expected to be very common in the LoTSS data at separations of less than 1.5\arcmin~\citep[see][fig.~2]{osullivan2020}. While these data products are useful for finding polarized sources and constructing an RM Grid, re-imaging would be required for detailed studies of the RM structure of individual sources. 

From an initial inspection of the source-finding results, it was noticed that the detections of many of the bright sources at low Faraday depth were actually most likely due to regions of instrumental polarization that extended in Faraday depth space outside the previously excluded leakage range. Therefore, an additional selection criterion was added to remove many of these sources. As noted above, the main peak of the FDF was also recorded, even if it occurred within the leakage range. Thus, in cases where the main peak occurred within the leakage range and the initially catalogued RM was within 3\rad~of this value, this source was then excluded from the catalogue. 

After these preliminary steps, there were 35161 candidate polarized source components identified. A conservative $8\sigma_{QU}$ threshold was then employed, which reduced the number of candidate polarized sources to 6744. This detection threshold was chosen in order to minimise the number of false detections in the final catalogue. For example, a false detection rate of $\sim10^{-4}$ is expected for an $8\sigma_{QU}$ threshold in the presence of non-Gaussian wings of the $Q$ and $U$ noise distribution, compared to as large as 4\% for a detection threshold of $5\sigma_{QU}$ \citep{george2012}. However, it is clear that many real sources exist below $8\sigma_{QU}$, and lowering this threshold can be one of the main ways to improve the areal number density of the LoTSS RM Grid in future work (Section~\ref{sec:future}). 

\subsubsection{Reliability of the detected polarized sources}\label{sec:fake}

After a detailed inspection of the preliminary catalogue of 6744 polarized source components, it was noticed that some fields had a much higher source density than expected (i.e.~$>50$ sources within 16~deg$^{2}$). One of these fields, P219+52, was particularly notable as it contained a 9~Jy source, 3C\,303, which was known to have a polarized intensity of $\sim$98~mJy~beam$^{-1}$ at 144 MHz \citep{vaneck2018}. However, in addition to 77 polarized sources identified above 8$\sigma_{QU}$ in this field, the polarized intensity of 3C\,303 was approximately half the expected value at only 49~mJy~beam$^{-1}$. 
The distribution of RM and fractional polarization for this field had a very narrow spread that was clearly unphysical (the median absolute deviations in the RM and the degree of polarization of the 77 sources in the field were 0.04\rad~and 0.05\%, respectively). Our conclusion was that approximately half the polarized flux of 3C\,303 had been transferred to other sources in the field. 

The most likely explanation for this behaviour was identified as the assumption that $Q = U = V = 0$~Jy, on average, for each LoTSS field in a direction-independent calibration step of the DR2 ddf-pipeline \citep{tasse2021}. 
The direction-independent full polarization calibration is done in at least 24 frequency bins across the HBA bandpass, and at a higher rate if the signal to noise ratio is high enough according to the criteria listed in section 3.1 of \cite{tasse2021}. 
This assumption has the advantage that it strongly suppresses the instrumental polarization (i.e.~leakage from Stokes $I$ into the other Stokes parameters). However, the major disadvantage was that for fields that were dominated by a bright, linearly polarized source (i.e.~$>10$~mJy~beam$^{-1}$ in polarized intensity), then spurious polarized sources were created throughout the field. These spurious or `fake' sources have very similar RM values to the brightest source in the field and have low fractional polarization values (i.e.~$\lesssim 0.5\%$). 

In order to understand the prevalence of this effect, we picked a LoTSS field with three weakly polarized sources, and injected a bright polarized source near the field centre into the {\sc prefactor} uv-data (using {\sc makesourcedb} and {\sc dppp}). The first test injected a 10 Jy point source that was 1\% linearly polarized. After running this field through the ddf-pipeline and RM synthesis, we found 75 fake polarized sources spread throughout the field. These fake sources had a narrow RM distribution centred on the input RM values, in addition to low fractional polarization values, similar to the 3C\,303 field. We then repeated this process several times, gradually reducing the polarized flux of the injected source, until no fake polarized source was detected in the field. This occurred at an input polarized flux of 15~mJy, where no fake polarized source was found in the field above a detection threshold of 7$\sigma_{QU}$. 

\subsubsection{Mitigation strategy}\label{sec:cuts}
We employed two main algorithms to identify and remove the fake polarized sources from the final catalogue. The first was a complete removal of polarized sources in particularly badly affected fields, and the second was based on identifying fields with a bright polarized source (defined as $p \geq 10$~mJy~beam$^{-1}$) and employing a more careful identification and removal of fake polarized sources. 

In order to identify the worst affected fields, we calculated the median absolute deviation (MAD) of the RM and the MAD degree of polarization of polarized sources for each field (see Fig.~\ref{fig:madRMmadfpol}, where each point is colour-coded by the number of polarized sources identified in a field). This allowed us to identify a region of parameter space where the fake polarized source problem was particularly severe (i.e.~the bottom left corner of Fig.~\ref{fig:madRMmadfpol}, where the MAD RM and MAD degree of polarization is very small and the number of polarized sources is unphysically large). 
The polarized sources in fields with MAD degree of polarization values greater than 10\% were also removed. These fields were dominated by imaging artefacts, making the identification of real polarized sources extremely challenging for the semi-automated procedure applied in this work. The fields within the shaded regions in Fig.~\ref{fig:madRMmadfpol} were excluded from the catalogue (corresponding to $\sim$12\% of the LoTSS-DR2 fields in total). 
Further work on identifying the problems in these fields is warranted but was deemed out of scope for the current work, where the main focus was on generating an initial, highly reliable catalogue of RMs from the LoTSS-DR2 data. 
We did not exclude all fields with a MAD degree of polarization less than 0.2\% (i.e.~the bottom right region of Fig.~\ref{fig:madRMmadfpol}), as these fields did not have large numbers of candidate polarized sources (unlike the fake source fields) and we considered it worthwhile to include this relatively small number of fields for more in-depth inspection. 

For the remaining fields, we identified sources with polarized intensities greater than 10~mJy~beam$^{-1}$ and then removed any other source in that field with an RM that is within 0.3\rad~of the RM of the bright source and with a degree of polarization ($p$) less than 1\% or greater than 20\%. Additionally, all sources with $p<0.2\%$ were removed in these fields. These criteria were defined from manual inspection of individual fields and from the injected source tests described above. It is possible that real polarized sources were removed using these criteria but we prefer to be conservative in this case to avoid any fake polarized sources remaining in the final catalogue. 

\begin{table*}
\caption{Sample columns from the LoTSS-DR2 RM Grid catalogue. The complete catalogue has 2461 rows and can be found through the links provided in Data Availability section. For the full list of catalogue columns and their descriptions, see Appendix~\ref{columns}. }\label{tab:RMall}
\begin{tabular}{cccccccccc}
\hline
RA        & Dec                           & RM   &           $|\bm{F}(\phi)|$                       &  $p$     &  $I$                    &  Source Name   & $z$ & ${\rm phot} = 0$ \\
  \[[J2000]      &     [J2000]        & [\rad] &  [mJy~beam$^{-1}$]  &   [\%]  &  [mJy~beam$^{-1}$]  &  LoTSS-DR2  &      &   ${\rm spec} = 1$  \\
\hline
0:01:32.6 & 24:02:33 & -66.406 $\pm$ 0.050 & 27.0 $\pm$ 0.3 & 1.84 $\pm$ 0.02 & 1469.4 $\pm$ 0.6 & ILTJ000132.27+240231.8 & 0.10448 & 1 \\
0:04:50.1 & 40:57:42 & -65.280 $\pm$ 0.062 & 1.5 $\pm$ 0.1 & 0.91 $\pm$ 0.06 & 164.8 $\pm$ 0.1 & ILTJ000451.64+405744.5 & -- & -- \\
0:05:07.5 & 40:57:06 & -63.086 $\pm$ 0.051 & 9.2 $\pm$ 0.1 & 4.49 $\pm$ 0.06 & 205.8 $\pm$ 0.1 & ILTJ000506.83+405711.8 & -- & -- \\
0:05:40.1 & 19:50:55 & -26.571 $\pm$ 0.050 & 28.6 $\pm$ 0.3 & 2.11 $\pm$ 0.02 & 1355.6 $\pm$ 0.7 & ILTJ000540.72+195022.4 & 0.6843 & 0 \\
0:05:59.4 & 35:02:02 & -64.643 $\pm$ 0.064 & 1.9 $\pm$ 0.1 & 1.34 $\pm$ 0.09 & 138.1 $\pm$ 0.2 & ILTJ000559.68+350204.4 & -- & -- \\
0:06:07.5 & 34:22:21 & -55.051 $\pm$ 0.056 & 1.9 $\pm$ 0.1 & 4.23 $\pm$ 0.18 & 45.3 $\pm$ 0.2 & ILTJ000607.41+342220.5 & 0.58427 & 1 \\
0:06:15.5 & 26:36:06 & -111.318 $\pm$ 0.051 & 7.0 $\pm$ 0.1 & 2.18 $\pm$ 0.03 & 324.0 $\pm$ 0.3 & ILTJ000624.41+263545.6 & 0.81131 & 0 \\
0:06:28.6 & 26:35:40 & -114.255 $\pm$ 0.052 & 2.9 $\pm$ 0.1 & 0.37 $\pm$ 0.01 & 773.7 $\pm$ 0.3 & ILTJ000624.41+263545.6 & 0.81131 & 0 \\
0:06:32.4 & 20:51:00 & -38.009 $\pm$ 0.063 & 1.4 $\pm$ 0.1 & 1.42 $\pm$ 0.09 & 100.7 $\pm$ 0.1 & ILTJ000632.39+205101.5 & 0.94204 & 0 \\
0:08:29.1 & 33:46:36 & -60.899 $\pm$ 0.051 & 5.4 $\pm$ 0.1 & 5.28 $\pm$ 0.09 & 103.1 $\pm$ 0.3 & ILTJ000828.85+334634.9 & -- & -- \\
0:08:32.1 & 42:17:50 & -49.277 $\pm$ 0.060 & 2.4 $\pm$ 0.1 & 0.79 $\pm$ 0.05 & 302.6 $\pm$ 0.2 & ILTJ000831.36+421725.0 & 1.0 & 0 \\
0:09:16.9 & 33:36:05 & -56.476 $\pm$ 0.058 & 1.7 $\pm$ 0.1 & 11.01 $\pm$ 0.56 & 15.0 $\pm$ 0.1 & ILTJ000916.76+333604.4 & -- & -- \\
0:10:04.3 & 30:45:45 & -63.880 $\pm$ 0.053 & 2.8 $\pm$ 0.1 & 0.88 $\pm$ 0.03 & 314.0 $\pm$ 0.2 & ILTJ001007.40+304524.3 & -- & -- \\
0:10:07.4 & 41:14:39 & -63.857 $\pm$ 0.051 & 9.0 $\pm$ 0.1 & 1.25 $\pm$ 0.02 & 721.4 $\pm$ 0.2 & ILTJ001006.00+411442.5 & -- & -- \\
. & . & . & . & . & . & . & . & . \\
. & . & . & . & . & . & . & . & . \\
. & . & . & . & . & . & . & . & . \\
23:59:51.9 & 39:40:54 & -117.322 $\pm$ 0.054 & 3.4 $\pm$ 0.1 & 2.70 $\pm$ 0.09 & 125.7 $\pm$ 0.2 & ILTJ235951.87+394052.7 & -- & -- \\
\hline
\end{tabular}
\end{table*}

\subsubsection{Further tests and future improvements}
Ongoing tests are being conducted to determine how to mitigate this problem for future datasets. The most obvious step is to remove the calibration step in the ddf-pipeline where the $Q = U = V = 0$~Jy assumption is made. This has already been implemented but has the downside that the instrumental polarization is no longer suppressed \citep[e.g.,][fig.~8]{tasse2021} and the fidelity of the final Stokes $I$ images is impacted. Alternatively, a polarized sky model could be developed of the $Q$ and $U$ flux in a field from the {\sc prefactor} data at low angular resolution, with this sky model then being incorporated into the ddf-pipeline. This approach is potentially quite expensive, as it requires RM synthesis and polarized source identification routines to be run in addition to the generation of $QU$ images cubes from the {\sc prefactor} data, before the ddf-pipeline is started. 
Presently, we are investigating the identification and subtraction of bright polarized sources from the {\sc prefactor} uv-data, before running this data through the ddf-pipeline and inspecting the output. Preliminary results are promising, but more work is needed. For example, an attempt to remove the source 3C\,303 from the P219+52 field (i.e.~subtracting a point source with a polarized intensity of $\sim$98~mJy using {\sc dppp} at the location of the peak polarized intensity), reduced the peak polarized intensity by $\sim$80\%, resulting in 45 fewer fake polarized sources above 8$\sigma_{QU}$ in that field (from 77 to 32). Clearly, further investigation of more comprehensive subtraction techniques are required. 
A different approach that could be applied to the current datasets is the subtraction of a normalised median FDF in $Q$ and $U$. This could work well in the fields that were completely discarded, where the problem is particularly severe, but may not be as effective in fields with moderate problems. 

\subsubsection{Final catalogue selection}
After addressing the fake polarized source issue, 4280 sources remained, for which some further automated cuts were made. Firstly, polarized sources that did not have a corresponding Stokes $I$ component in the LoTSS catalogue were removed. Secondly, an additional leakage source removal step was employed for sources with $|{\rm RM}| < 5$\rad~and $p < 1$\%, where the source was removed only if its RM value deviated from the MAD RM for that field by more than 10\rad~(i.e.~sources that deviated significantly from the mean RM of the field and had a low degree of polarization were plausibly leakage and thus removed to ensure the high reliability of the catalogue). 
This leakage source removal step is an extension of the previous leakage exclusion range in Faraday depth of $-3$\rad~to $+1$\rad~for all sources (Section~\ref{sec:original}), and of $\pm3$\rad~of the Faraday depth value of the main leakage peak in the FDF (Section~\ref{sec:leakage2}). Example FDFs for the various types of excluded sources are shown in Fig.~\ref{fig:leakage}. 

This catalogue was then cross-matched with itself (within 1\arcmin) in order to identify the duplicate sources due to the large overlap between fields. There were 1381 duplicates found, with the largest number of duplicates for any one source being 4. The source that was closest to the nearest field centre was retained. These duplicate sources are an excellent means of assessing the systematic error in the RM values (see Section~\ref{sec:rmgrid}). This left 2559 unique source components, of which cutout images were prepared for a final quality assessment by visual inspection. These cutout images consisted of plots of the absolute value of the FDF, the polarized intensity image, the RM image and the degree of polarization image (all overlaid by Stokes $I$ contours), in addition to the NVSS total intensity and polarized intensity image at 1.4~GHz. 
The visual inspection led to the removal of a small number of sources, in addition to identifying candidate pulsars (see Section~\ref{sec:pulsars}). 

For each entry in the final LoTSS RM Grid catalogue of 2461 polarized components, we extracted the single-pixel $Q$ and $U$ versus frequency spectra at the component location. 
While {\sc pyrmsynth} was the most efficient RM synthesis software for running on the large cubes (with the 1 mJy~beam$^{-1}$ Stokes $I$ threshold), we switched to using the {\sc rm-tools} package \citep{RMtools}\footnote{https://github.com/CIRADA-Tools/RM-Tools} to analyse the final selected polarized sources more comprehensively. 
Thus, we re-ran RM synthesis on the extracted spectra with {\sc rm-tools}, using the same Faraday depth range of $\pm120$\rad, but with a higher sampling in Faraday depth of 0.05\rad~and weighting each channel by the inverse variance of the channel noise. The catalogued RM and polarized intensity was obtained by fitting a parabola to the main peak outside of the leakage range as defined above, and correcting for polarization bias following \citet{george2012}. The catalogue output columns follow the RMTable standardised format\footnote{https://github.com/CIRADA-Tools/RMTable}, as described in detail in Van Eck et al.~in prep. %\cite{vaneck2022}. 
The details for how to access the final catalogue and the associated advanced data products are provided in the Data Availability section. 
Descriptions of the catalogue columns are provided in Appendix~\ref{columns}, with a selection of columns shown in Table~\ref{tab:RMall}. 

\begin{figure*}
\includegraphics[width=17.5cm,clip=true,trim=0.0cm 0.0cm 0.0cm 0.0cm]{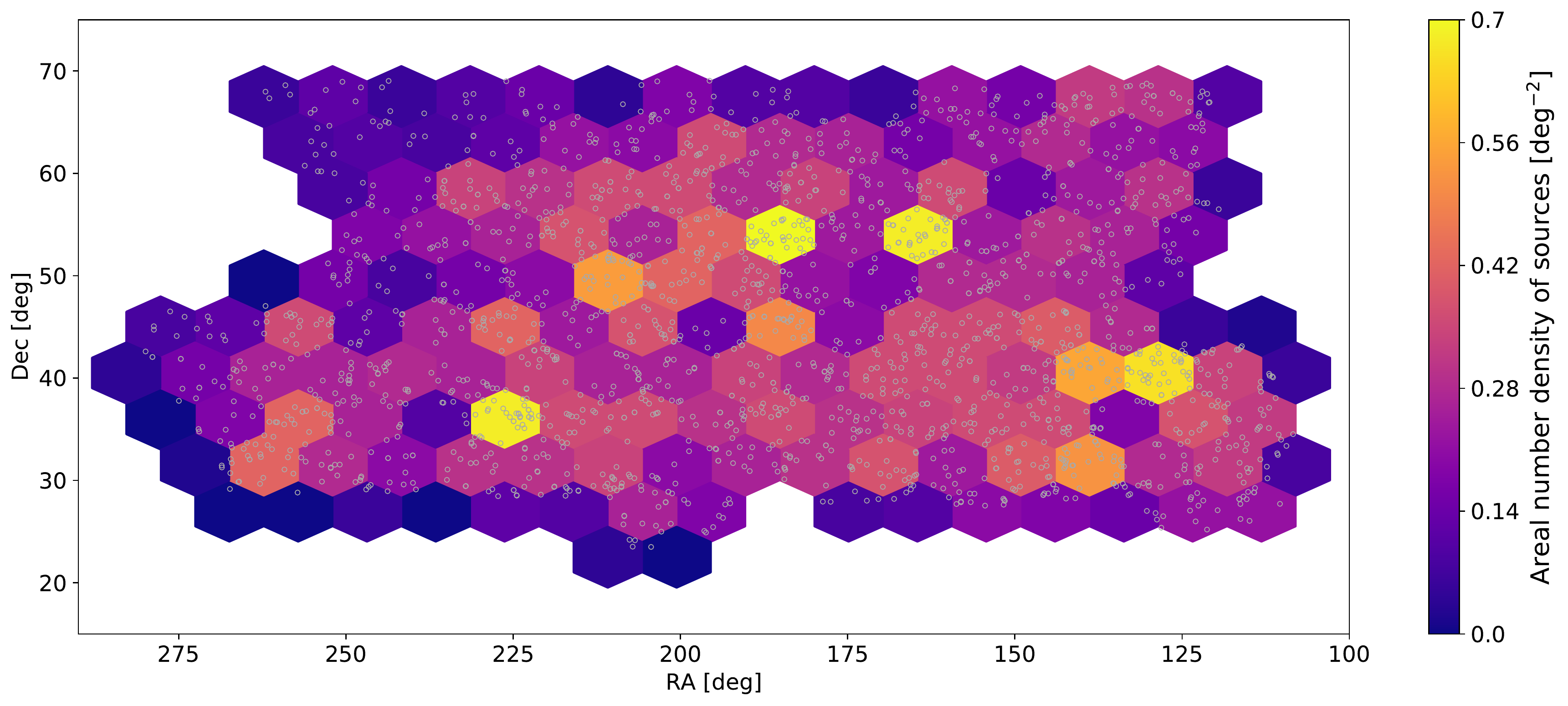}
\caption{ The polarized source areal number density across the 13~hr field, shown as a 2-D histogram with hexagonal cells, with the colour scale in units of deg$^{-2}$. The locations of the polarized sources are shown by open grey circles. The polarized source areal number density reaches values as high as 0.7~deg$^{-2}$, with the average value being 0.48~deg$^{-2}$. }\label{fig:3b}
\end{figure*}

\section{Results}\label{sec:results}

\subsection{LoTSS-DR2 RM Grid}\label{sec:rmgrid}
The sky distributions of the 2461 source components in the RM Grid are shown in Fig.~\ref{fig:10} for the 13~hr field and Fig.~\ref{fig:11} for the 0~hr field, in orthographic projection, with the red/blue circles denoting positive/negative RM values and the size of the circles being proportional to the magnitude of the RM. The large coherent patches of positive and negative RM values highlight how the RM contribution from the Milky Way (i.e.~the Galactic RM or GRM) dominates the mean RM values. An all-sky Hammer-Aitoff projection of the RM Grid in celestial coordinates is shown in Fig.~\ref{fig:12}, with both the 13~hr and 0~hr fields visible, along with grey points indicating the locus of the Galactic plane. Fig.~\ref{fig:13} shows the coverage of the LoTSS RM Grid in Galactic coordinates and highlights the complementarity of the POGS RM survey \citep{riseley2020}, which has a lower polarized source density, using the Murchison Widefield Array (MWA) in the Southern Hemisphere. There is a small area of overlap between the two surveys around $l=120$\degr, $b=-40$\degr~which is investigated in Section~\ref{sec:pogs}. 

\begin{figure}
\includegraphics[width=8.5cm,clip=true,trim=0.0cm 0.0cm 0.0cm 0.0cm]{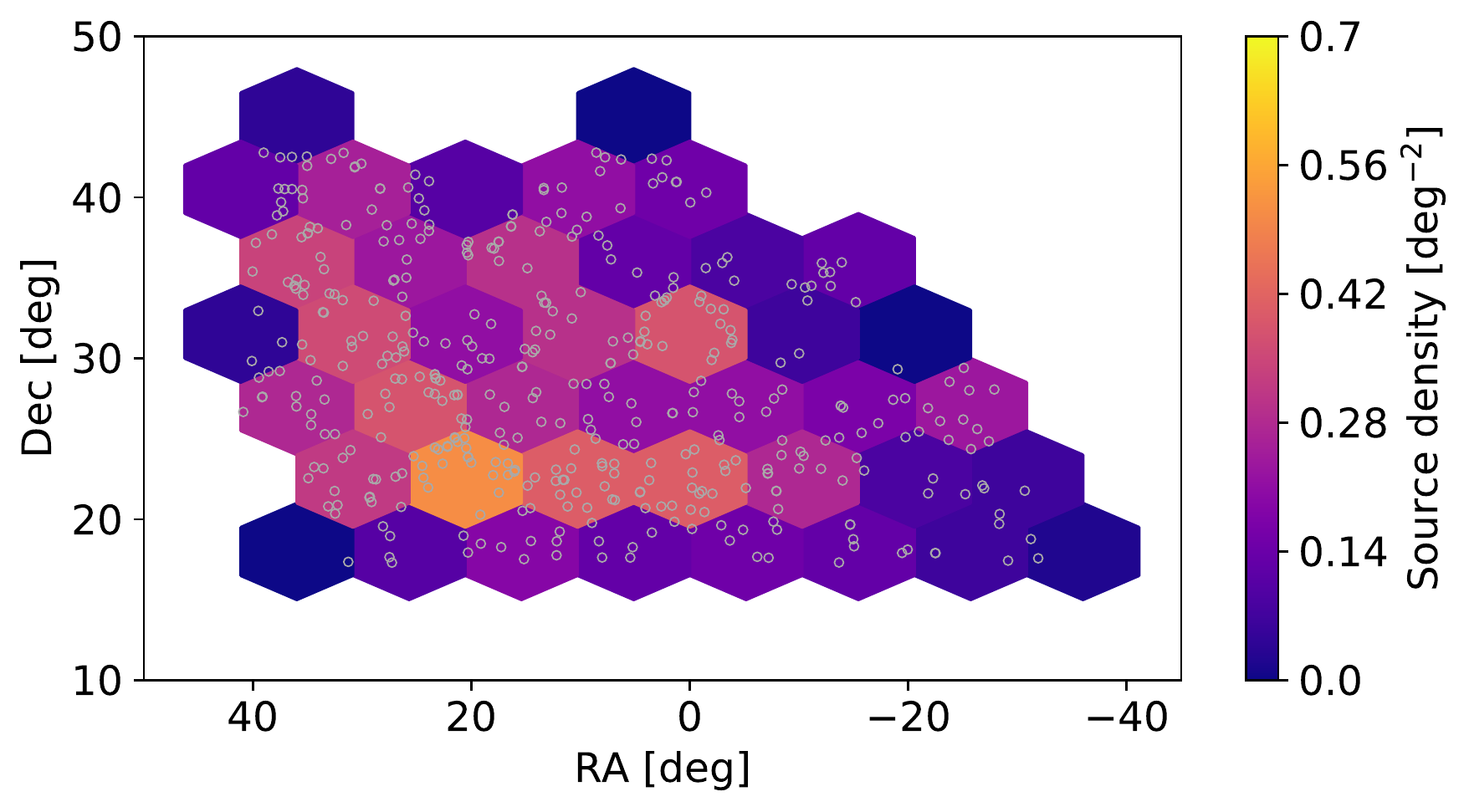}
\caption{ The polarized source areal number across the 0~hr field, with the same colour-scale as in Fig.~\ref{fig:3b} for direct comparison (it shows a lower areal number density in general, with an average of 0.29~deg$^{-2}$). }\label{fig:3c}
\end{figure}

There are 2,039 RM Grid sources (82\%) in the 13~hr field, and 422 in the 0~hr field. Therefore, the areal number density on the sky is 0.48~deg$^{-2}$ in the 13~hr field and 0.29~deg$^{-2}$ in the 0~hr field (the areal number density variations across the fields are shown in Figures~\ref{fig:3b} and~\ref{fig:3c}). The 0~hr field covers a region with 
negative GRM values ranging from $-218$ to $-40$\rad~\citep{oppermann2012}. Therefore, the 
lower number density is possibly due to some missing high $|$RM$|$ sources that were outside the Faraday depth range of $\pm120$\rad~that was searched, in addition to the slightly worse sensitivity due to the lower Declination of the field. 

\begin{figure}
\includegraphics[width=8.5cm,clip=true,trim=0.0cm 0.25cm 0.0cm 0.0cm]{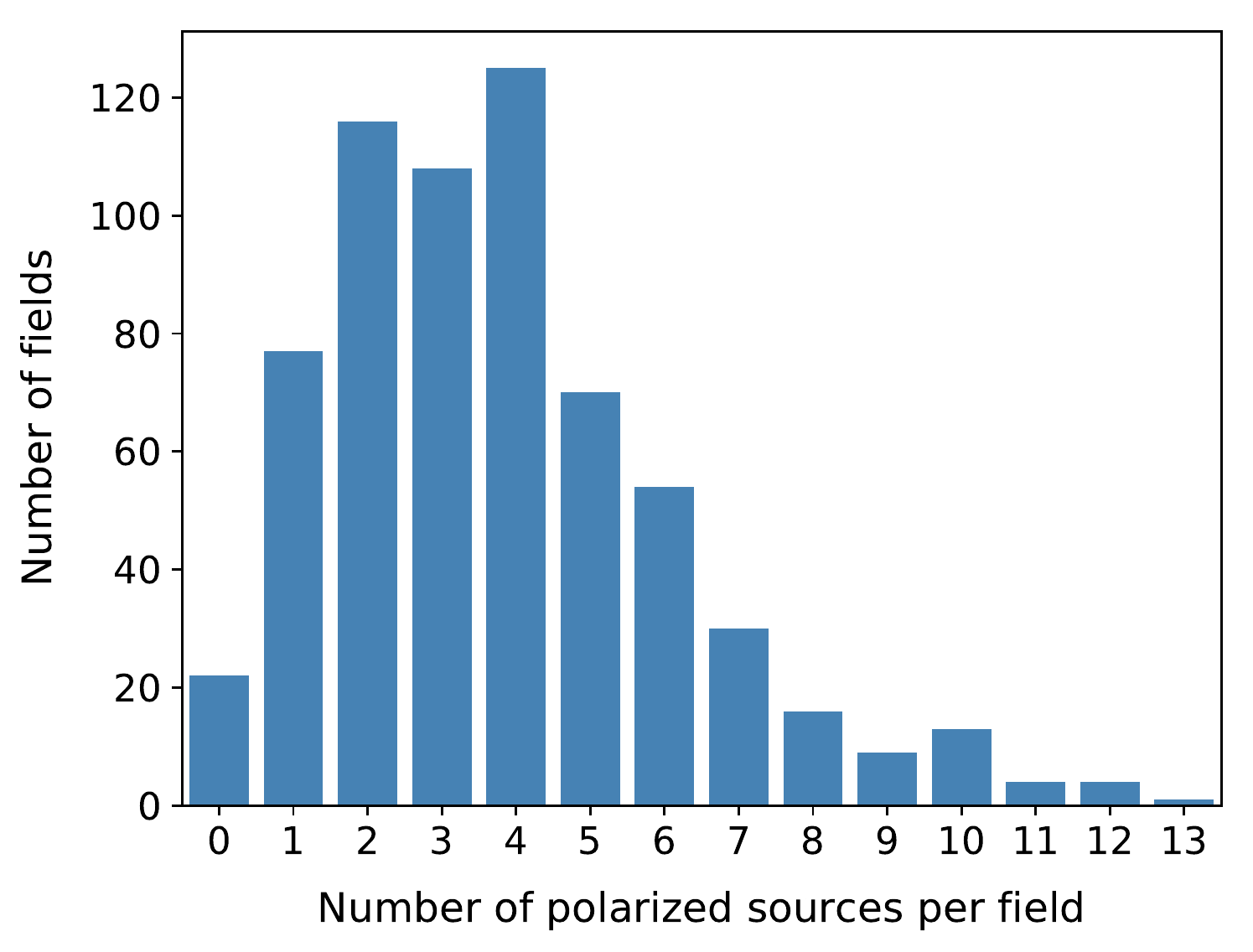}
\caption{Histogram of the number of polarized sources per field (i.e.~over an area of 16~deg$^{2}$). The median number of polarized sources per field is 4. }\label{fig:3a}
\end{figure}

We note that 173 fields were removed from the analysis (out of a total of 844 fields) by the data quality cuts described in Section~\ref{sec:cuts}, even though, as noted, real polarized sources are likely to be found in these fields with more advanced algorithms. Therefore, the true polarized source areal number density achievable with the LoTSS data at 20\arcsec~is expected to be larger than that presented here. Also, future analysis of the polarization data at 6\arcsec~resolution is expected to reveal more polarized sources. Of the fields that were analysed, only 22 had zero polarized sources detected. The median number of polarized sources per field was 4 and the maximum was 13. Figure~\ref{fig:3a} shows a histogram of the number of polarized sources per field. 

\begin{figure}
\includegraphics[width=8.5cm,clip=true,trim=0.0cm 0.25cm 0.0cm 0.0cm]{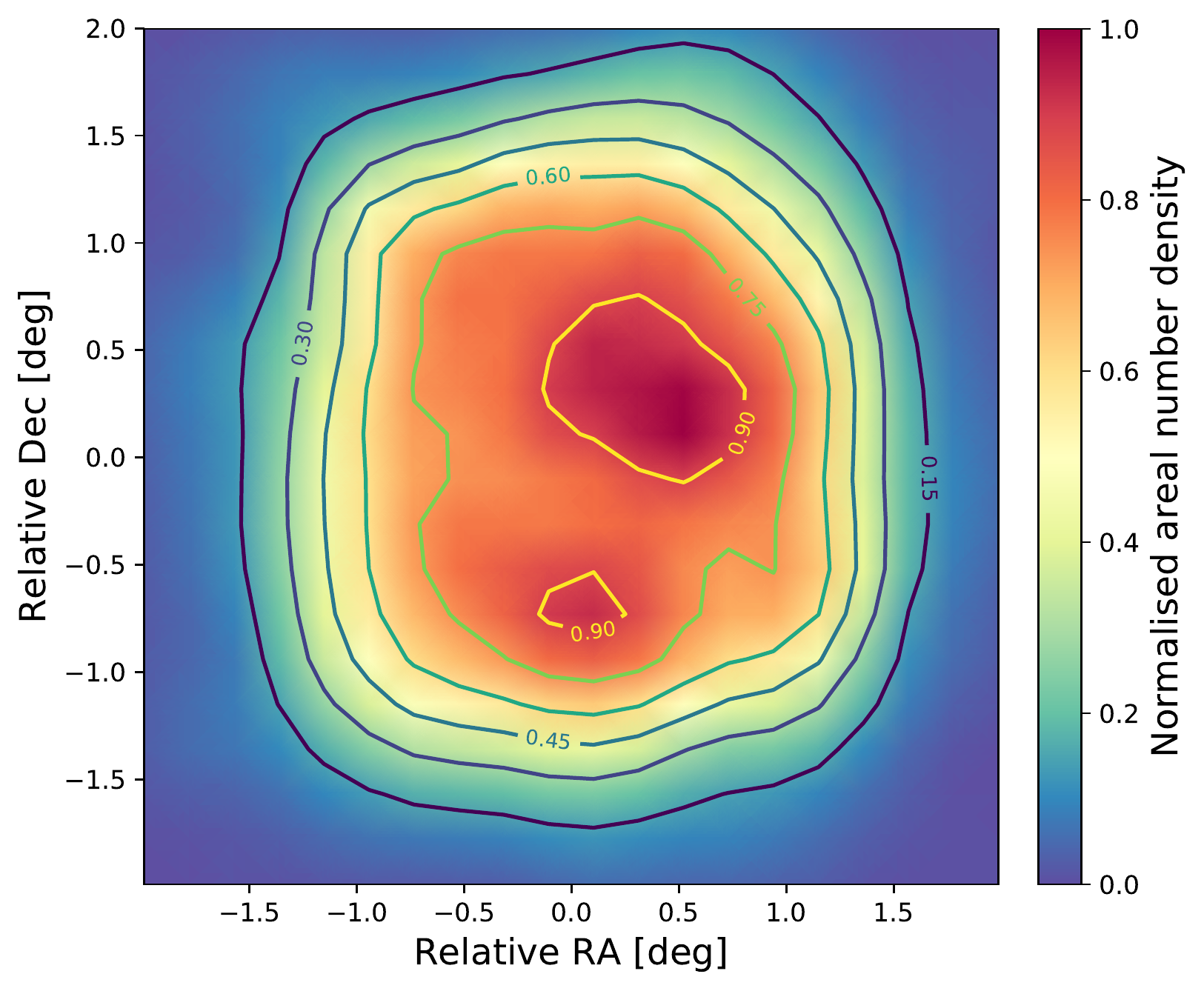}
\caption{ The normalised areal number density of polarized sources, averaged over all LoTSS-DR2 fields for the 16~deg$^{2}$ areas analysed. This shows that more sources are found near the centre of the fields, where the noise is lower, as expected. }\label{fig:9}
\end{figure}

The $Q$ and $U$ images were not mosaicked (due to the absence of an absolute polarization angle calibration), so the noise is non-uniform across the DR2 area (the median noise in the Faraday spectra of the detected sources is 0.08~mJy~beam$^{-1}$). Fig.~\ref{fig:9} shows the normalised source number density averaged over all fields, showing that most sources are detected within the central parts of the field with the lowest noise, as expected. 

\begin{figure}
\includegraphics[width=8.5cm,clip=true,trim=0.0cm 0.0cm 0.0cm 0.0cm]{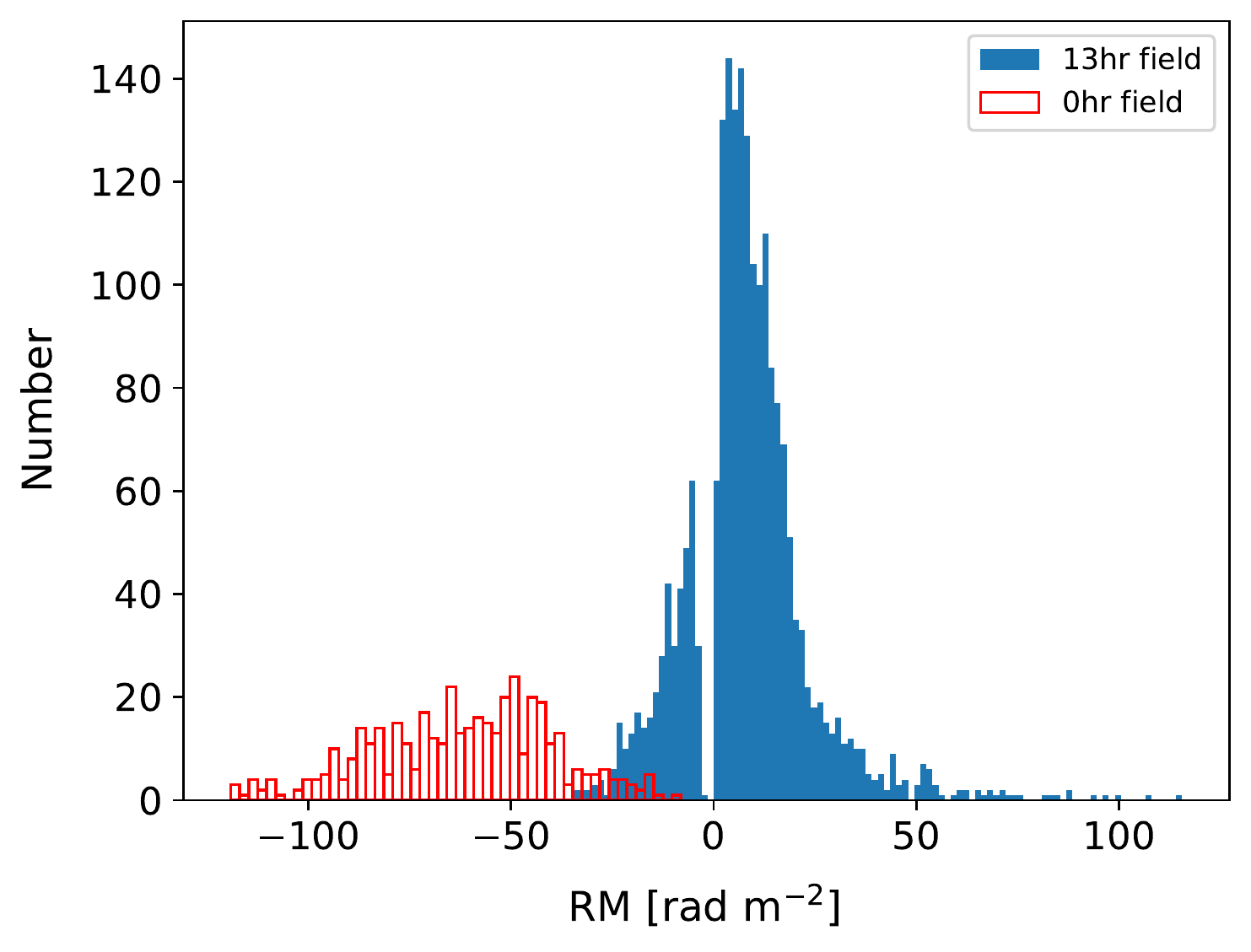}
\includegraphics[width=8.5cm,clip=true,trim=0.0cm 0.25cm 0.0cm 0.0cm]{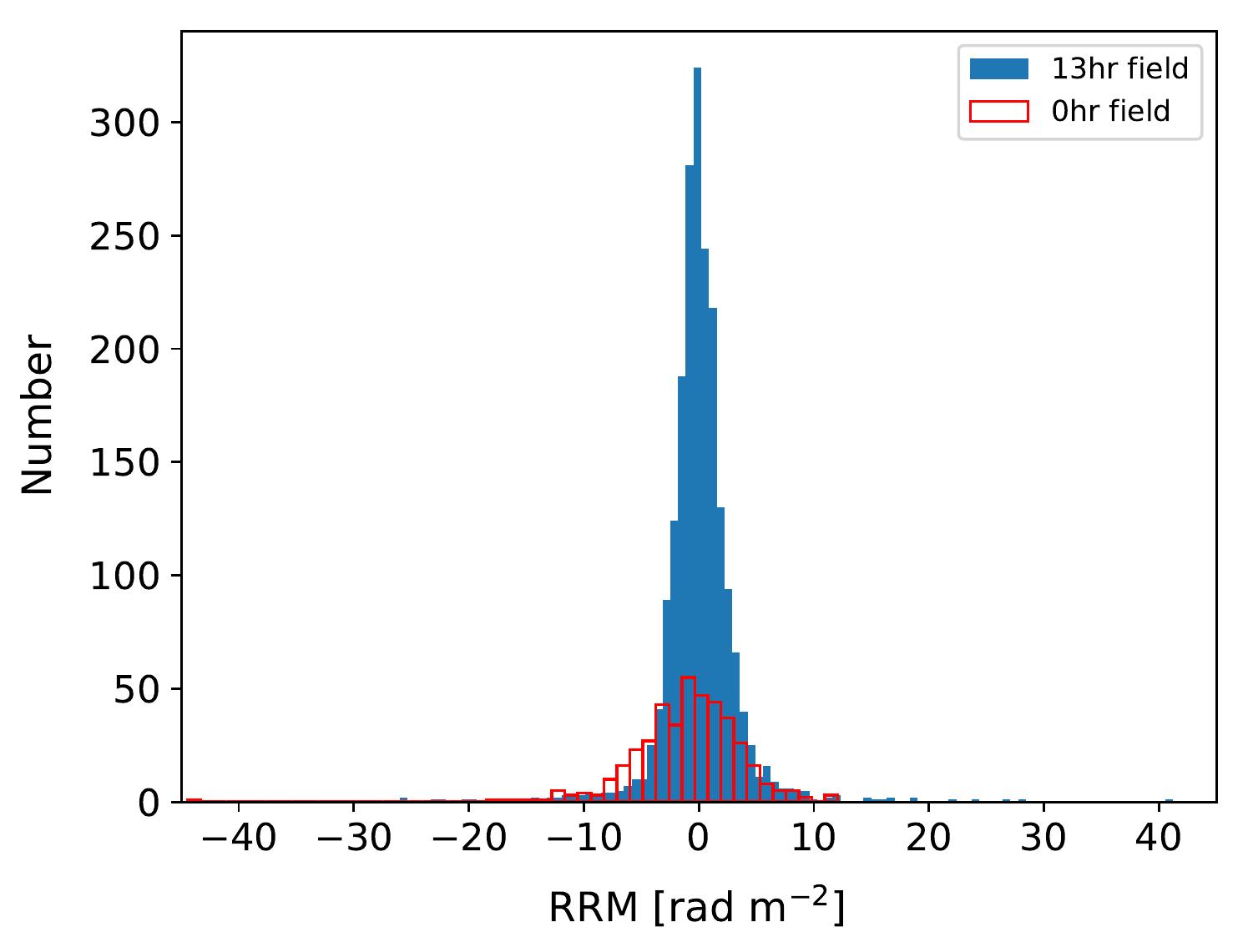}
\caption{ Top: histogram of the RM values for the 2461 polarized sources in both the 13~hr field and 0~hr field of the LoTSS-DR2 area. The spread represents the dominant contribution of the Galactic RM. The gap near 0\rad~is due to the exclusion of sources contaminated by the instrumental polarization present in LOFAR data. 
Bottom: histogram of the residual rotation measure (RRM) after subtraction of the model Galactic RM (GRM), for the 13~hr field (blue solid) and for the 0~hr field (red open). 
}\label{fig:4}
\end{figure}

\begin{figure}
\includegraphics[width=8.5cm,clip=true,trim=0.0cm 0.0cm 0.0cm 0.0cm]{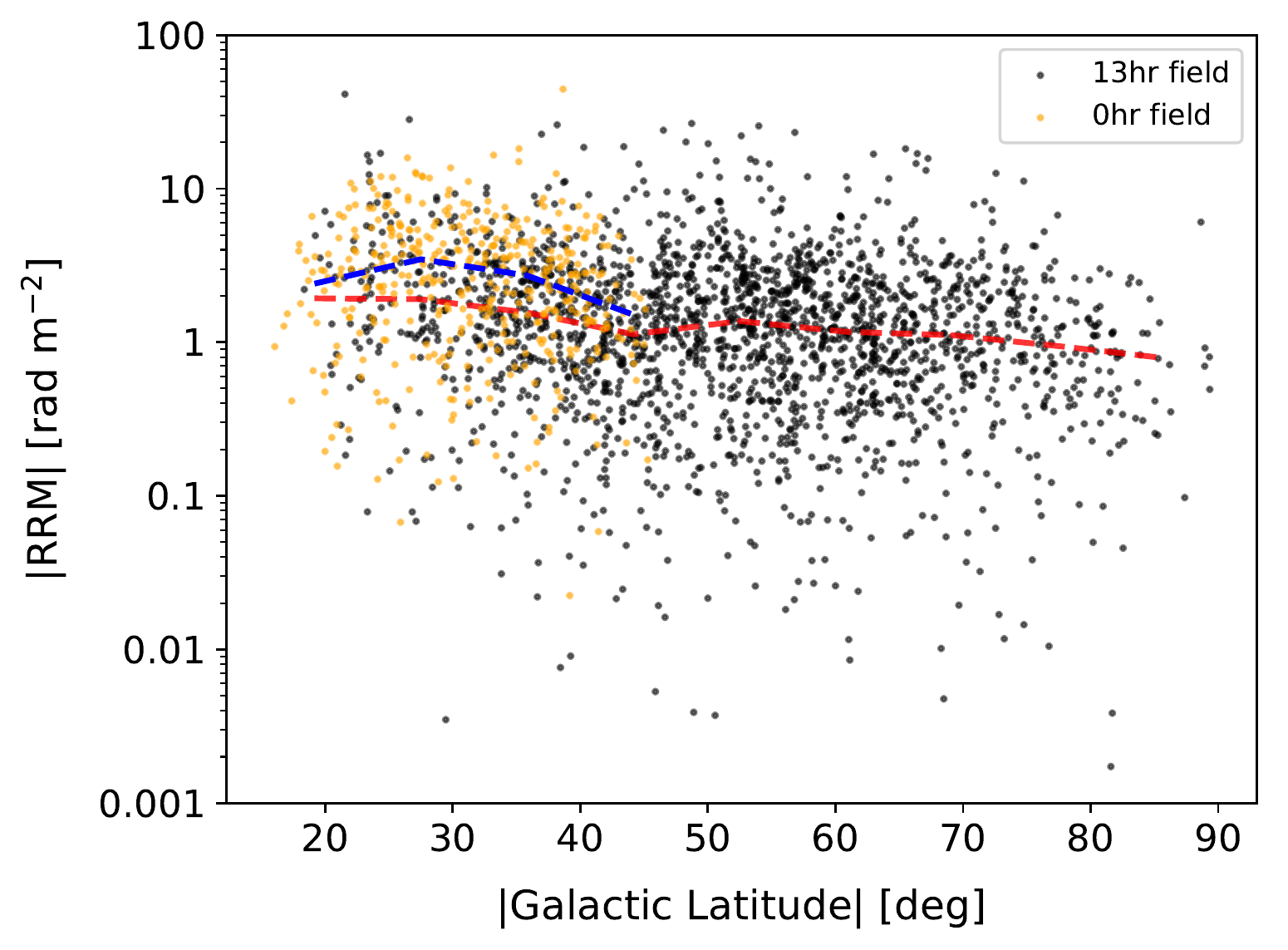}
\caption{ The absolute value of the residual RM, $|$RRM$|$, as a function of the Galactic latitude for the 13~hr field (black points) and for the 0~hr field (orange points). The red(blue) dashed line shows the running median of the 13~hr(0~hr) field, which highlights a slight trend that remains in the RRM after subtraction of the Galactic RM model, mainly for absolute Galactic latitudes less than 45 degrees. }\label{fig:6}
\end{figure}

The distribution of the observed RM values is shown in Fig.~\ref{fig:4},~top, where it is clear that RM values near 0\rad~are missing from the data (due to the leakage exclusion range employed in this work). Since the RM values are dominated by the GRM, we also show the residual RM (RRM) after subtraction of the GRM in Fig.~\ref{fig:4},~bottom (i.e.~${\rm RRM}= {\rm RM} - {\rm GRM}$). Here we used the GRM model from \citep{hutsch2022}, and subtracted the average GRM value within a 1 degree diameter disc surrounding each RM position. A 1 degree diameter is chosen because this is the typical separation between the input data points in the \citet{hutsch2022} model (see also \citet{carretti2022} who used a similar approach). 
The mean of the RRM distribution is $-0.15$\rad, and the robust standard deviation (excluding outliers) is $\sim$2\rad, which is a combination of the real extragalactic RM variance and the uncertainty in the GRM model values. The 13~hr field has a smaller RRM standard deviation of $\sim$1.8\rad, compared to the 0~hr field standard deviation of $\sim$4\rad. A slight trend in the $|$RRM$|$ versus Galactic latitude 
is evident (Fig.~\ref{fig:6}), suggesting that improved GRM models and/or subtraction techniques remain desirable. 

\begin{figure*}
\includegraphics[width=17.5cm,clip=true,trim=0.0cm 0.0cm 0.0cm 0.0cm]{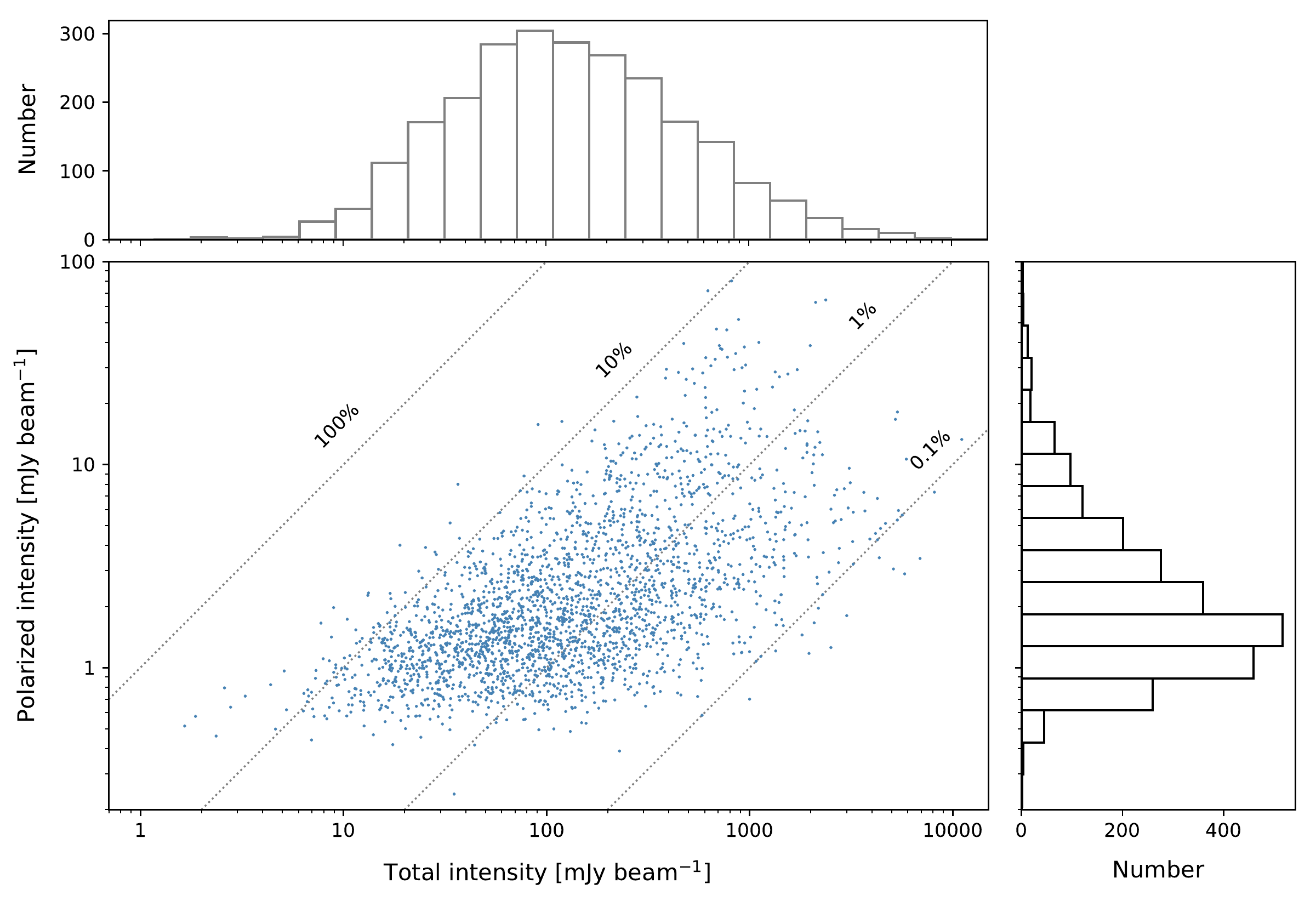}
\caption{ Distribution of the peak linearly polarized intensity versus the total intensity, in units of mJy~beam$^{-1}$, for the 2461 polarized sources in the LoTSS-DR2 RM Grid. The diagonal dashed lines represent constant values of the degree of polarization.  }\label{fig:1}
\end{figure*}

In Fig.~\ref{fig:1} we show the total intensity and polarized intensity of the 2461 detected sources, with the range of the degree of polarization indicated by the diagonal lines. The median polarized intensity is 1.7~mJy~beam$^{-1}$, while the median total intensity is $\sim$120~mJy~beam$^{-1}$. There are $\sim$1.2 million catalogued total intensity sources in the LoTSS-DR2 area with peak flux density brighter than 1~mJy~beam$^{-1}$, meaning that only $\sim$0.2\% of sources are detected in polarization above a threshold of 8$\sigma_{QU}$ (i.e.~$\sim$0.6~mJy~beam$^{-1}$). 

\begin{figure}
\includegraphics[width=8.2cm,clip=true,trim=0.0cm 0.0cm 0.0cm 0.0cm]{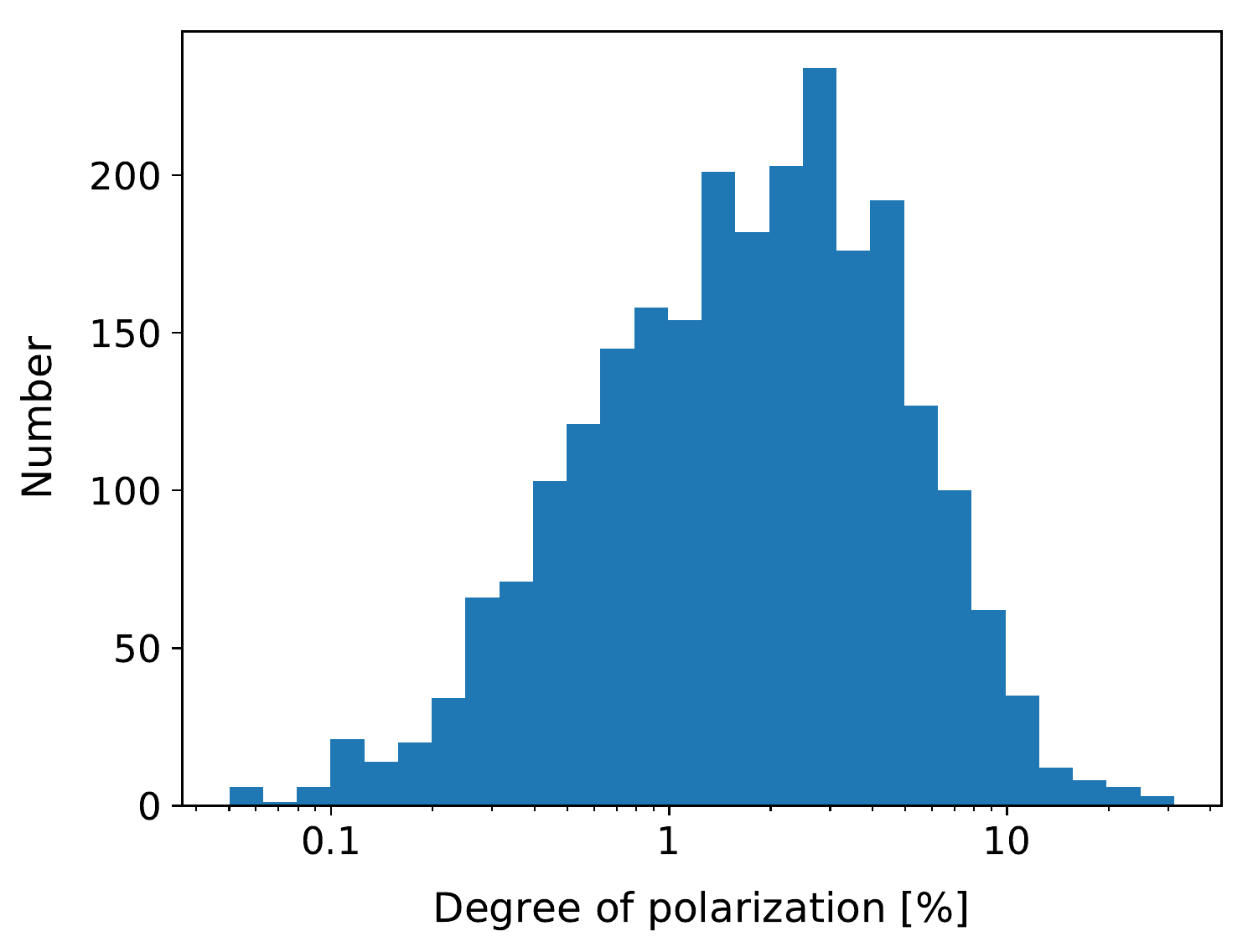}
\caption{ Histogram of the degree of polarization of the LoTSS-DR2 polarized sources, ranging from 0.05\% to 31\% with a median value of 1.8\%. }\label{fig:fpol}
\end{figure}

\begin{figure}
\includegraphics[width=8.2cm,clip=true,trim=0.0cm 0.2cm 0.0cm 0.0cm]{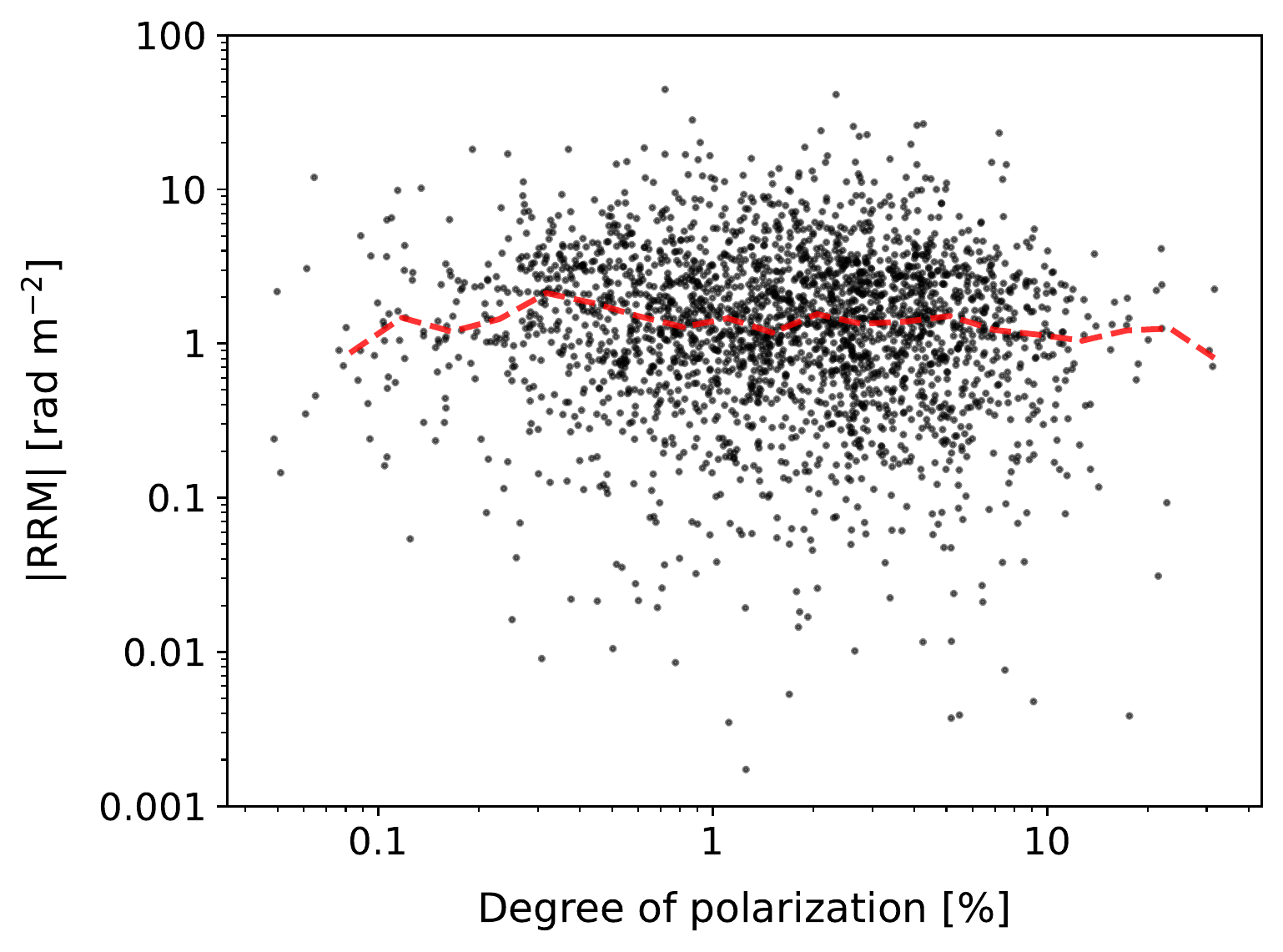}
\caption{ The absolute value of the residual RM ($|$RRM$|$), after subtraction of the model Galactic RM, versus the LoTSS degree of polarization. The red dashed line shows the running median of the $|$RRM$|$ for equally spaced bins in log-space. }\label{fig:RRMfpol}
\end{figure}

The median degree of polarization of the detected sources is 1.8\%, ranging from 0.05\% to 31\% (see Fig.~\ref{fig:fpol} for a histogram of the degree of polarization). 
It is notable that very low degrees of polarization ($\lesssim0.1\%$) are detectable due to the narrow RMSF of LoTSS ($\sim$1.16\rad), because the instrumental polarization peak around 0\rad~typically only contaminates a small region of Faraday depth space around this value (depending on how bright the instrumental polarization peak is for any particular source). The LoTSS degree of polarization appear to be independent of the RRM (Fig.~\ref{fig:RRMfpol}), which is in contrast to other studies at cm-wavelengths \citep[e.g.][]{hammond2012}. As argued in \cite{carretti2022}, this potentially implies that the LoTSS RRM is not dominated by contributions local to the source but instead from the intergalactic medium. However, detailed depolarization studies of the LoTSS sources are required to more reliably isolate the local source effects \citep[e.g.][]{stuardi2020}.

\begin{figure}
\includegraphics[width=8.5cm,clip=true,trim=0.0cm 0.0cm 0.0cm 0.0cm]{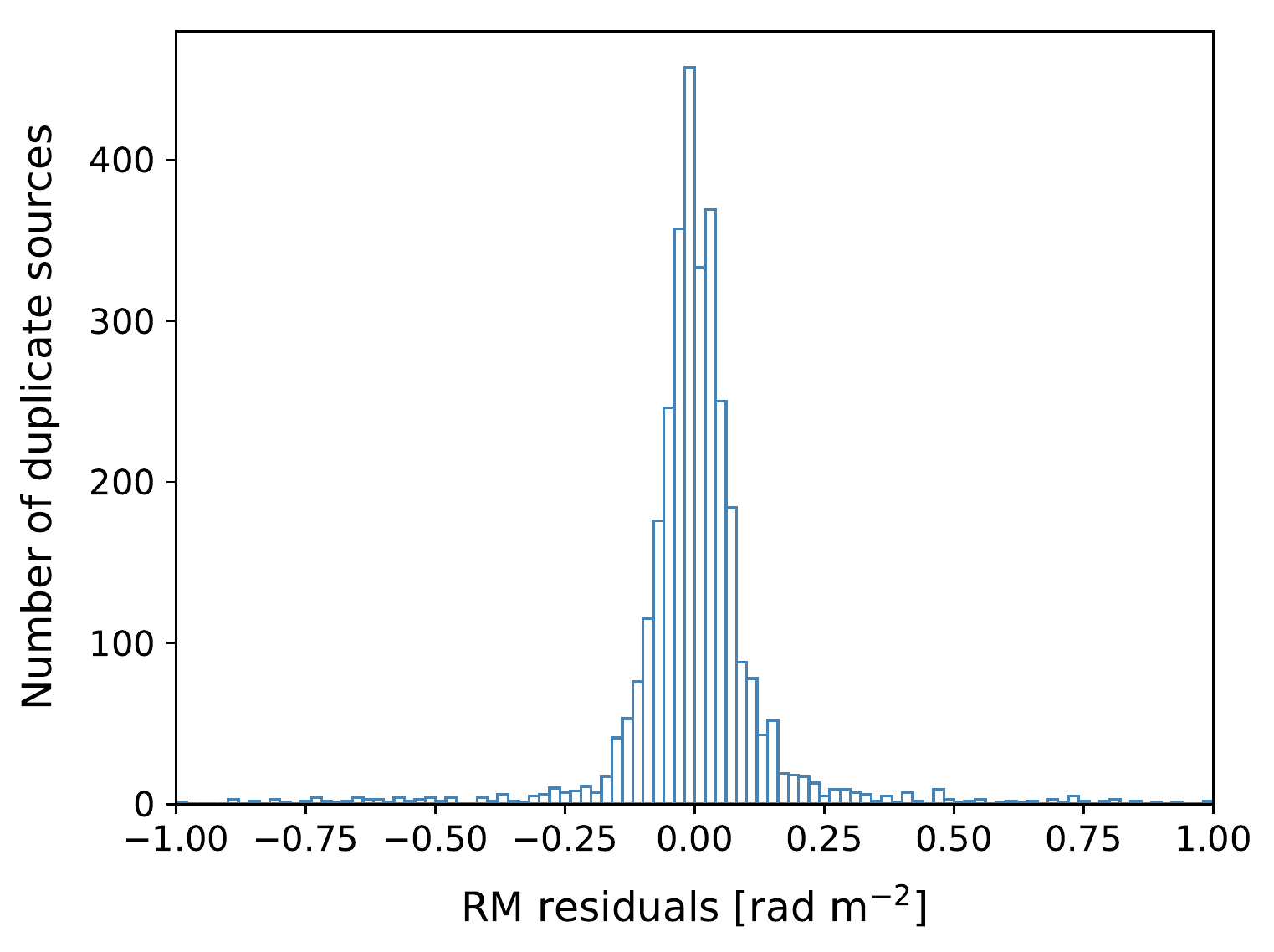}
\caption{ Histogram of the difference from the average RM for sources detected multiple times (i.e.~in more than one field). }\label{fig:RMerr}
\end{figure}

Due to the large overlap between adjacent LoTSS fields, there were 1,380 duplicate RM measurements (918/365/97 sources in two/three/four different fields). The variation in RM between the multiple observations provides a means to assess the systematic error in the LoTSS RM values, most likely due to ionosphere RM correction errors. Fig.~\ref{fig:RMerr} shows the difference from the average RM values for these duplicate sources, 
with a robust standard deviation of 0.067\rad~(i.e.~calculated as 1.4826 times the median absolute deviation). Part of the spread is due to the standard measurement error in the RM estimate from the signal to noise (i.e.~the FWHM of the RMSF divided by twice the signal to noise), so subtracting this contribution in quadrature leaves a systematic error estimate of $\sim$0.05\rad. Therefore, in the catalogue, we list two different errors in the RM values, one including the systematic error estimate of 0.05\rad~in the RM between fields, and one based solely on the signal to noise (which would be relevant for an analysis of the RM difference of close pairs within the same field, for example). 

\subsection{Comparison with other RM catalogues}
\subsubsection{NVSS RM catalogue}
The NRAO VLA Sky Survey (NVSS) RM catalogue at 1.4~GHz \citep{taylor2009} overlaps with the entire LoTSS area and so is an excellent resource for checking the LoTSS data reliability as well as for depolarization studies \citep[e.g.][]{stuardi2020}. In general, LoTSS detects much fewer polarized sources than the NVSS in a given sky area, due to the much stronger effect of Faraday depolarization at 144~MHz compared to 1.4~GHz \citep[e.g.][]{sokoloff1998}. However, of the 2461 LoTSS polarized sources, only 910 (37\%) are also in the NVSS RM catalogue. Therefore, the majority of LoTSS sources provide unique RM values, which is important for RM Grid studies that want to maximise the areal number density of RM values on the sky (irrespective of at what frequency they were determined). The reason there are LoTSS polarized sources that are not detected in the NVSS is because the LoTSS survey is $\sim$10 times more sensitive for steep spectrum radio sources \citep[e.g.][]{osullivan2018b,mahatma2021}, coupled with the three times higher angular resolution which helps to reduce the effect of beam depolarization in general. 

\begin{figure}
\includegraphics[width=8.5cm,clip=true,trim=0.0cm 0.0cm 0.0cm 0.0cm]{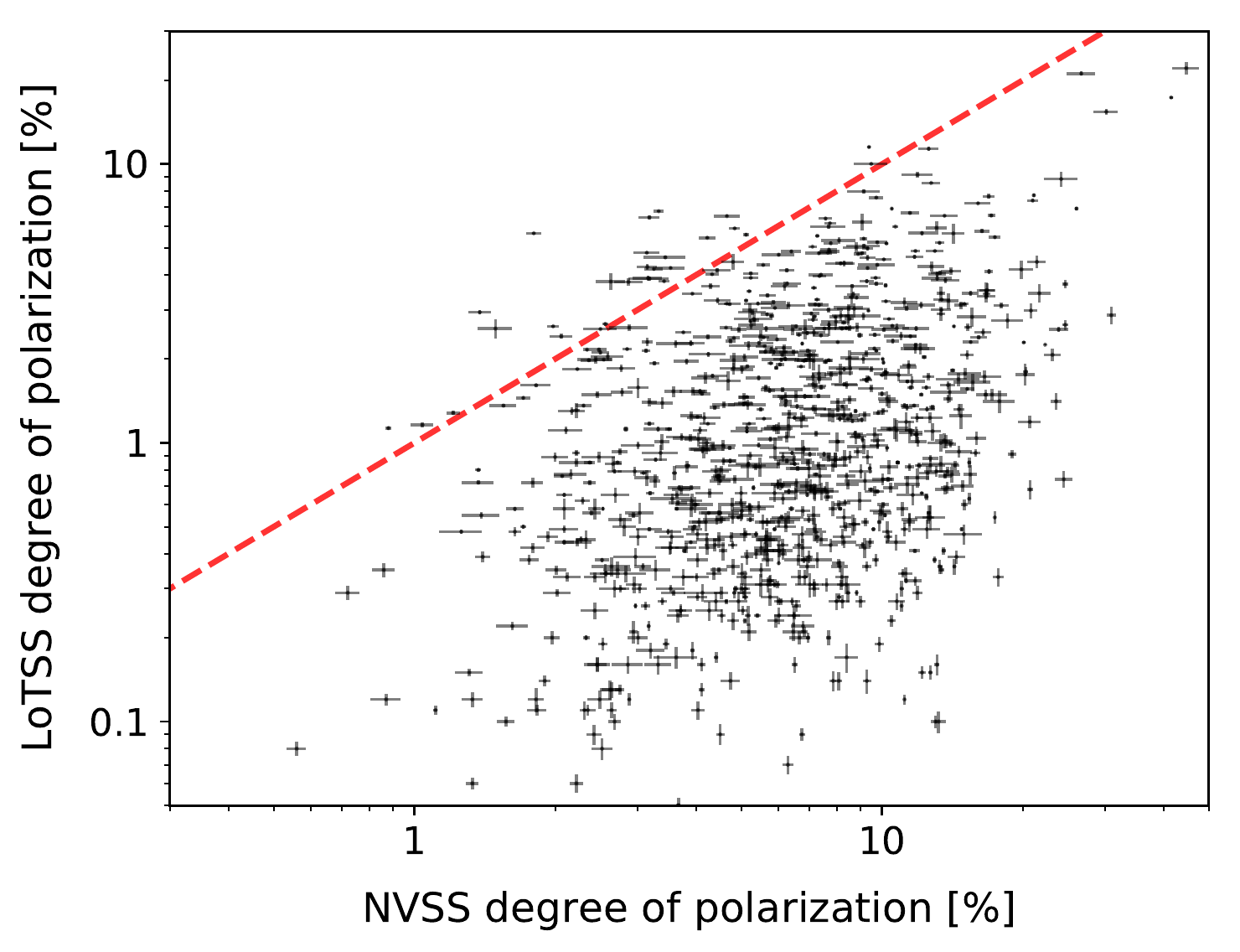}
\caption{ The degree of polarization of the LoTSS-DR2 sources (at 144~MHz) compared to the corresponding NVSS (at 1.4~GHz) catalogued polarized sources (within 1\arcmin). The red dashed line represents the one-to-one correspondence, which highlights how the majority of the sources in common are strongly affected by Faraday depolarization at 144 MHz. Note that the LoTSS $QU$ data has an angular resolution of 20\arcsec~while the NVSS RM catalogue was created from 60\arcsec~images \citep{taylor2009}. }\label{fig:7}
\end{figure}

For those sources in common, Fig.~\ref{fig:7} shows the degree of polarization comparison, where the dashed red line represents the one-to-one relation. This shows that almost all the sources in common have a higher degree of polarization in the NVSS compared to LoTSS, which is as expected due to the increased effect of Faraday depolarization at low frequencies. The few sources that have a higher degree of polarization in LoTSS could be explained by either intrinsic source variability (e.g.~blazars) or the higher angular resolution of LoTSS for sources that experience only very small amounts of Faraday depolarization. 
We note that the median degree of polarization of all sources in the NVSS RM catalogue is 5.8\%, however, the sources in common with LoTSS have a median degree of polarization of 6.6\%. This highlights how LoTSS detections are preferentially selecting for sources with high degrees of polarization at 1.4 GHz (i.e.~low depolarization or `Faraday simple' sources). 

\begin{figure}
\includegraphics[width=8.0cm,clip=true,trim=0.0cm 0.0cm 0.0cm 0.0cm]{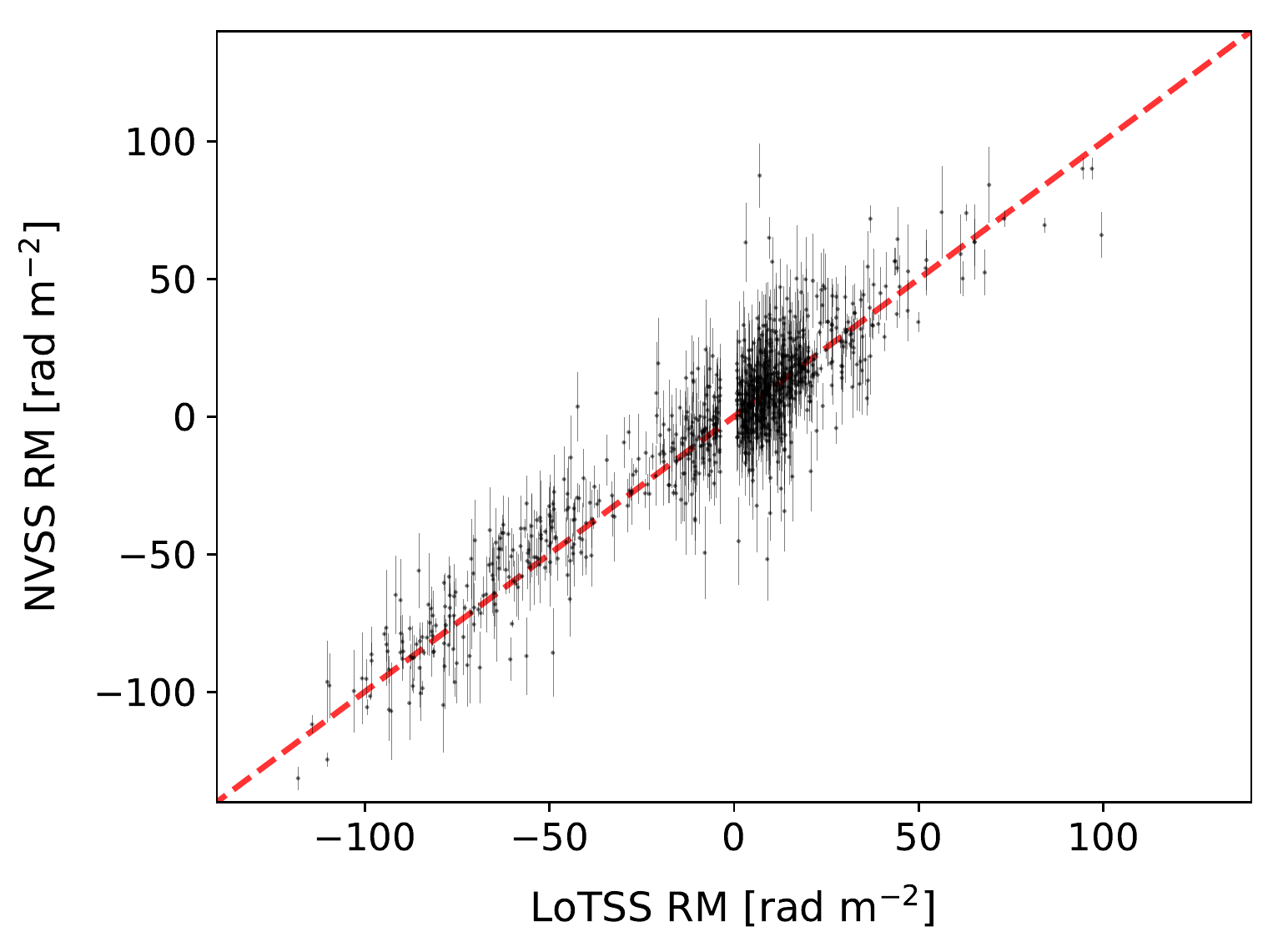}
\includegraphics[width=8.8cm,clip=true,trim=0.0cm 0.0cm 0.3cm 1cm]{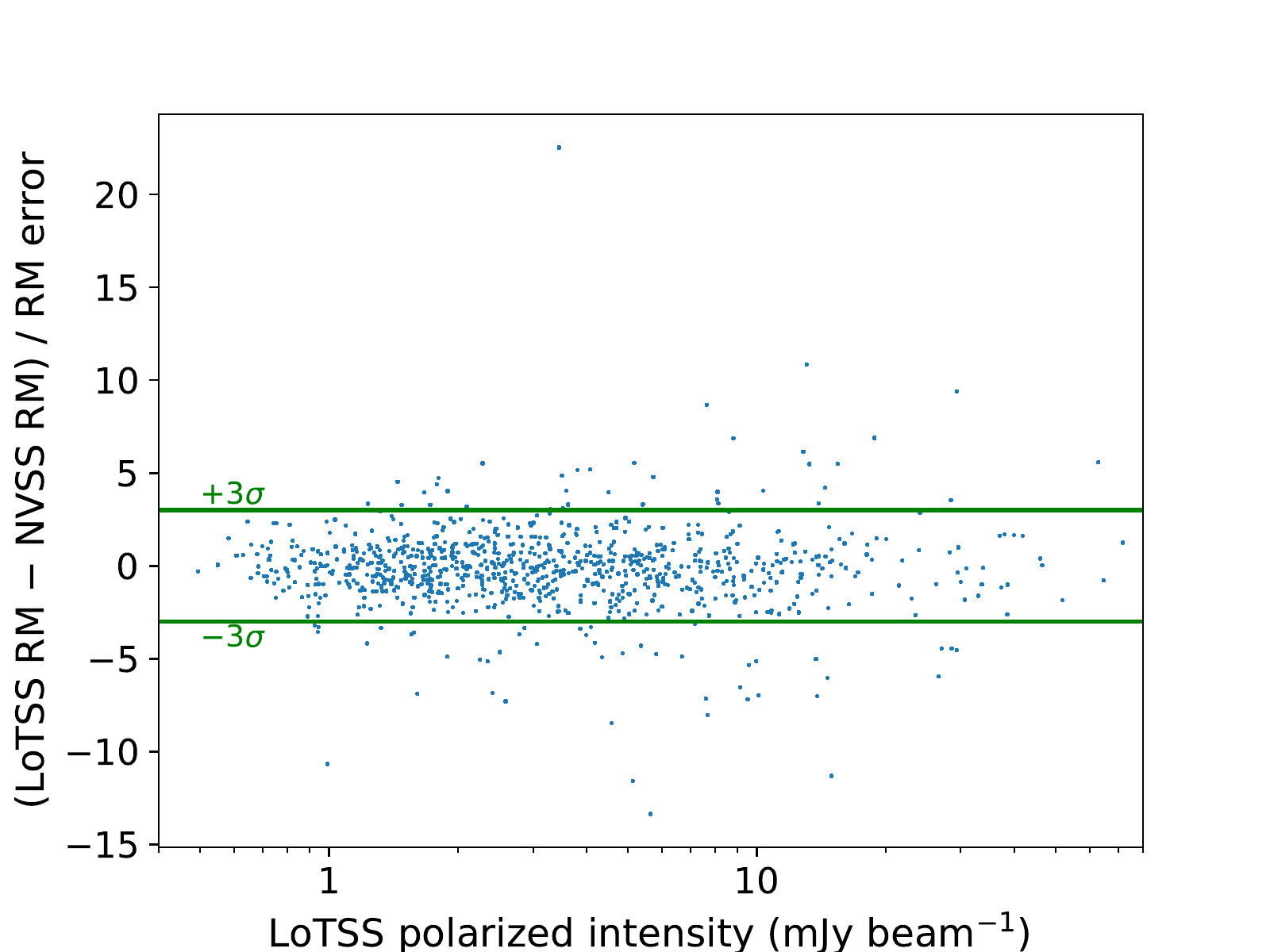}
\caption{ Top: The direct comparison of the NVSS RM values with the corresponding LoTSS RM values (within 1\arcmin). The red dashed line represents the one-to-one correspondence, which highlights the general good agreement considering the large NVSS RM errors compared to LoTSS RM errors (not visible on the plot). Lower: The difference between the LoTSS RM and NVSS RM values divided by the combined RM error, as a function of the LoTSS polarized intensity. This shows that 90\% of the RM values agree within $3\sigma$ (solid green lines). One possibility for the outliers would be intrinsic source variability. }\label{fig:8}
\end{figure}

In general, the NVSS and LoTSS RM values agree within the uncertainties, as shown by the direct comparison between the RM values in Fig.~\ref{fig:8},~top, where the dashed red line represents the one-to-one relation. More quantitatively, Fig.~\ref{fig:8},~bottom shows the RM difference relative to the combined RM error (which is dominated by the NVSS RM errors) versus the LoTSS polarized intensity values. This shows that 90\% of sources agree within 3$\sigma$ (solid green lines) and also that there are no systematic differences between the NVSS and LoTSS RM values as a function of the LoTSS polarized intensity. This highlights that there is minimal Faraday complexity in the polarized emission of sources detected by the LoTSS survey, which makes them excellent sources for RM Grid studies. 
A more detailed investigation of the outliers in Fig.~\ref{fig:8} would be worthwhile, as they could possibly be explained by the more robust RM determination of LoTSS (i.e.~much better wavelength-squared coverage) compared to the NVSS \citep{ma2019}, the detection of different polarized regions 
of the source with different RM properties within the synthesised beams, or due to intrinsic source variability (e.g.~blazars). 

\subsubsection{Other cm-wavelength RM catalogues}
A recent compilation by Van Eck et al.~in preparation\footnote{https://github.com/CIRADA-Tools/RMTable}, 
includes a number of other datasets with RMs derived at cm-wavelengths. We find 66 sources in common (that are not from the NVSS RM catalogue or other m-wavelength catalogues such as MWA-POGS described below).  Of these 66 sources, 71\%(83\%) of the RM values agree within $3\sigma$($5\sigma$) of the combined errors. Thus, these are slightly more discrepant than the general NVSS population. However, the majority of these discrepant RM values are from the \cite{farnes2014a} catalogue which has a large number of flat spectrum sources and are thus likely to display intrinsic RM variability and/or significant Faraday complexity \citep[e.g.][]{anderson2019}. 

We also compare our results with recent work at $\sim$1.4~GHz described in \cite{adebahr2022}, where they derive robust RM values from polarized sources in the Apertif Science Verification Campaign (SVC), which covers 56~deg$^2$ in five non-contiguous fields. 
Three of the SVC fields (containing 901 polarized sources) overlap with the LoTSS RM Grid catalogue coverage, two with the LoTSS 13~hr field (containing 593 SVC sources) and the other with the 0~hr field. In the 13~hr(0~hr) field there are 11(3) polarized sources in common. This means that there are 64 times more polarized sources found in the SVC compared to LoTSS in the overlap regions in total, and 54 times if we just compare to the 13~hr field regions. Given that the two surveys have similar sensitivities to typical steep spectrum sources, this is representative of the general expectations since we expect $\sim$53 times more sources in the SVC than in LoTSS because $\sim$10.57\%($\sim$0.2\%) of Stokes $I$ sources are found to be polarized in the SVC(LoTSS). Directly comparing the RM values of the sources in common, we find that 3 out of 14 are different by more than five times the combined errors. On closer inspection, 2 of the 3 discrepant RMs are actually from the opposite lobe of the same source, while the other discrepant RM comes from a BL Lac object, so variability and/or Faraday complexity is likely responsible for this difference. 

\subsubsection{MWA-POGS RM catalogue}\label{sec:pogs}
Part of the DR2 0~hr field (south of Dec.~$+30$\degr) overlaps with the POSG-II RM catalogue from \cite{riseley2020}, which used the Murchison Widefield Array (MWA) from 169 to 231~MHz. By cross-matching the two RM catalogue positions within 3\arcmin~we find 6 sources in common. The RM and degree of polarization values are shown in Table~\ref{tab:pogs} for comparison. 
The six LoTSS RM values in common are consistently more negative than the POGS values, by an average of 0.6\rad. Considering the different angular resolutions (20\arcsec~vs. 3\arcmin) and frequencies (144 MHz vs.~200 MHz), this difference may not be so surprising, however the systematic offset suggests this is worth further investigation (e.g.~differences in the ionosphere RM corrections). In any case, this comparison provides another estimate of the true uncertainty in these RM values at low frequencies. 

There are another 5 POGS-II polarized sources that are within the DR2 area but not present in the LoTSS RM catalogue (POGS-II-033, -036, -068, -441, -442). As the degrees of polarization for these sources are all below 3\%, this absence of these sources may be due to Faraday depolarization. However, the RM values for these sources ($-8.2$\rad, 19.3\rad, 10.9\rad, 19.4\rad, 19.4\rad, respectively) are quite different from the typical negative RM values in this region (e.g.~Table~\ref{tab:pogs}), which may indicate the presence of Faraday complexity (for which POGS is more sensitive given the higher observing frequency). 

\begin{table*}
\centering
\caption{Comparison of RM values with sources in common with the POGS-II catalogue \citep{riseley2020}.}
\label{tab:pogs}
\begin{tabular}{lccccccccc}
\hline
Source Name & $p_{\rm LoTSS}$ & RM$_{\rm LoTSS}$ & POGS ID & RM$_{\rm POGS}$ & $p_{\rm POGS}$ & RM$_{\rm POGS}$$ - $RM$_{\rm LoTSS}$ & Separation  \\
LoTSS-DR2 & [\%] & [\rad] & & [\rad] & [\%] &  [\rad] & [\arcsec] \\
\hline
ILTJ000540.72+195022.4 & $2.11\pm0.02$ & $-26.571\pm0.050$    & POGSII-EG-007 & $-25.768\pm0.077$ & $2.5\pm0.4$ & 0.8  & 5.1\\
ILTJ010118.71+203129.7 & $1.67\pm0.04$ & $-27.901\pm0.052$       & POGSII-EG-047 & $-27.402\pm0.128$ & $7.3\pm1.9$  & 0.5 & 30.5\\
ILTJ014751.99+223855.4 & $5.33\pm0.06$ & $-39.046\pm0.050$   & POGSII-EG-071 & $-38.922\pm4.259$ & $3.7\pm1.3$  & 0.1 & 36.4\\
ILTJ230010.12+184537.5 & $3.89\pm0.05$ & $-64.484\pm0.050$    & POGSII-EG-462 & $-63.700\pm0.168$    & $15.9\pm3.2$ & 0.8 & 21.9\\
ILTJ233518.42+174026.5 & $4.50\pm0.05$ & $-45.119\pm0.050$    & POGSII-EG-475 & $-44.613\pm0.121$ & $4.4\pm0.7$ &  0.5 & 17.9\\
ILTJ235945.26+203610.1 & $1.55\pm0.03$ & $-42.512\pm0.051$     & POGSII-EG-484 & $-41.292\pm7.791$ & $2.9\pm1.9$  & 1.2 & 74.3\\
  \hline
\end{tabular} 
\end{table*}

\subsection{Optical identification and redshifts}
We conducted a LOFAR Galaxy Zoo effort \citep[e.g.][]{williams2019} within the Surveys and Magnetism Key Science Project teams just for the polarized sources (which we label as the MKSP-LGZ) in order to (i)~associate each polarized source with the correct total intensity source, and (ii)~identify the host galaxy. The MKSP-LGZ used the LoTSS catalogued Stokes $I$ components and contours, overlaid on image panels with LEGACY optical \citep{dey2019} and WISE infrared \citep{wright2010} images, in addition to a cross symbol which marked the location of the peak polarized emission. 
Each source was subjected to five classification attempts by astronomers, which resulted in the association of all polarized source components with a unique LoTSS-DR2 source name (e.g.~ILTJ000132.27+240231.8) and the host galaxy identification for 2168 of the 2461 polarized components (88\%). 
We note that there are 17 catalogue entries which do not have an ILT source name, as they lie just outside the edge of the catalogued LoTSS-DR2 area in Stokes $I$ (which only includes sources above the 0.3 power point of the primary beam).  

\begin{figure}
\includegraphics[width=8.5cm,clip=true,trim=0.0cm 0.2cm 0.0cm 0.0cm]{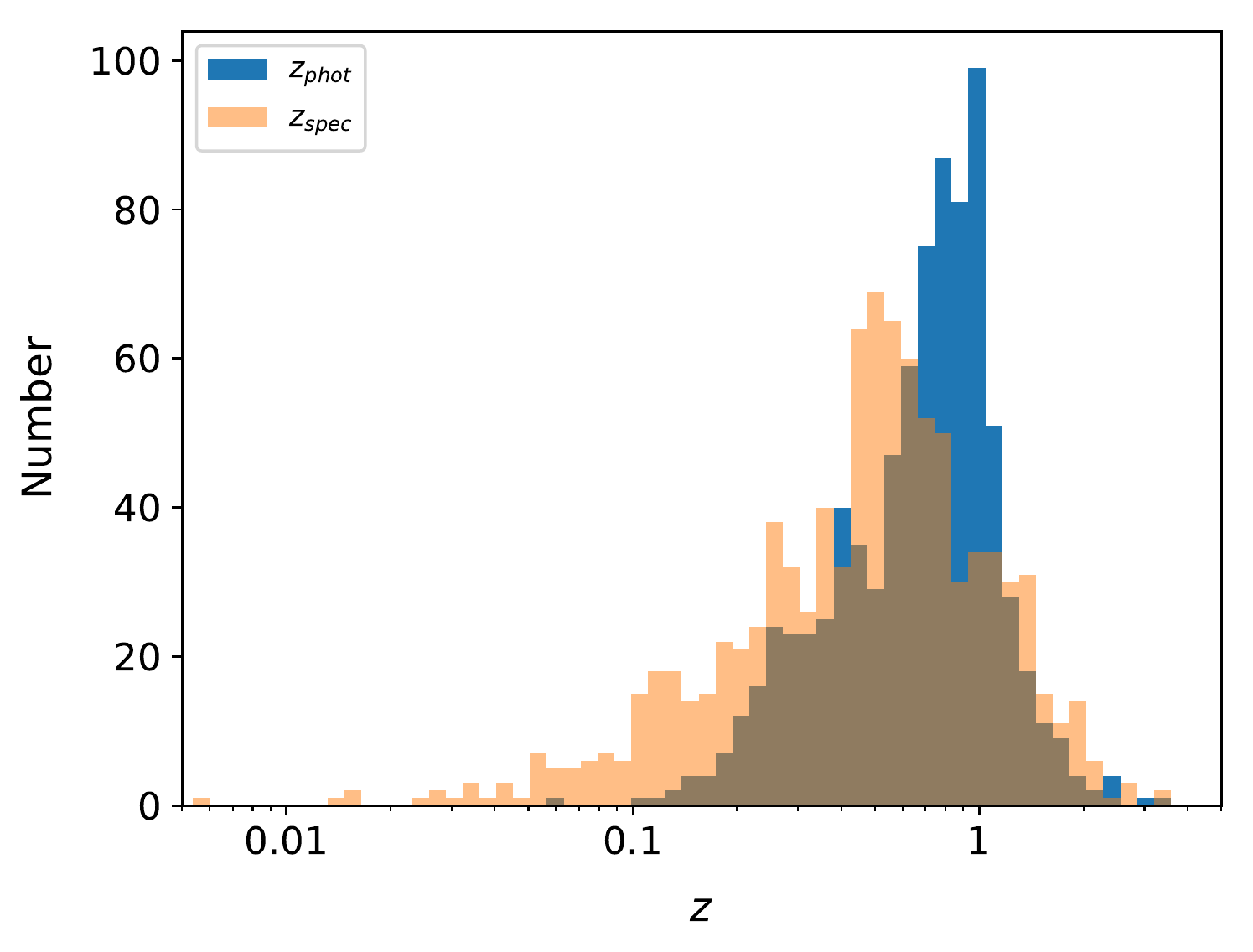}
\caption{ Histogram of the spectroscopic (orange) and photometric (blue) redshift values for the host galaxy identification of the LoTSS-DR2 polarized sources ($\sim$79\% of the sample has either a photometric or spectroscopic redshift). }\label{fig:z}
\end{figure}

Photometric redshifts for 1641 of the host galaxies were obtained from a hybrid template fitting and machine learning approach \citep{duncan2021}. Further work enabled us to find 938 spectroscopic redshifts for the host galaxies in the literature (associated with 1046 RM entries in the catalogue, meaning some sources have RM entries for both lobes). 
From this we define a ``z\_best'' column which 
contains 1949 entries corresponding to each RM value with an associated redshift. Because some sources have multiple RM entries (e.g.~physical pairs), in total there are 1762 unique source redshifts, where spectroscopic redshifts supersede photometric ones, leaving 824 phot-$z$ in addition to 938 spec-$z$. Therefore, 79\% of the RM Grid catalogue entries have associated redshifts. 
The distributions of spectroscopic and photometric redshifts are show in Fig.~\ref{fig:z}. The median for all redshifts is $z_{\rm med}=0.6$, while $z_{\rm med,spec}=0.5$ and $z_{\rm med,phot}=0.7$. 

Many studies have investigated the dependence of the extragalactic RM on redshift \citep[e.g.][]{orenwolfe1995, hammond2012, xuhan2022}. The high precision in the RM values from LoTSS, in addition to the high fraction of sources with redshifts, provides an excellent opportunity for new discoveries. 
The data presented here show a flat behaviour of the RRM versus redshift and a general decrease in the degree of polarization with redshift by a factor of $\sim$10 between $z\sim0$ and $z\sim3$. We refer the reader to \cite{carretti2022,carretti2022b} for detailed investigations of the astrophysical implications of these behaviours. 

\subsubsection{Radio luminosity, linear size \& morphology}\label{sec:morph}
\begin{figure}
\includegraphics[width=8.5cm,clip=true,trim=0.0cm 0.0cm 0.0cm 0.0cm]{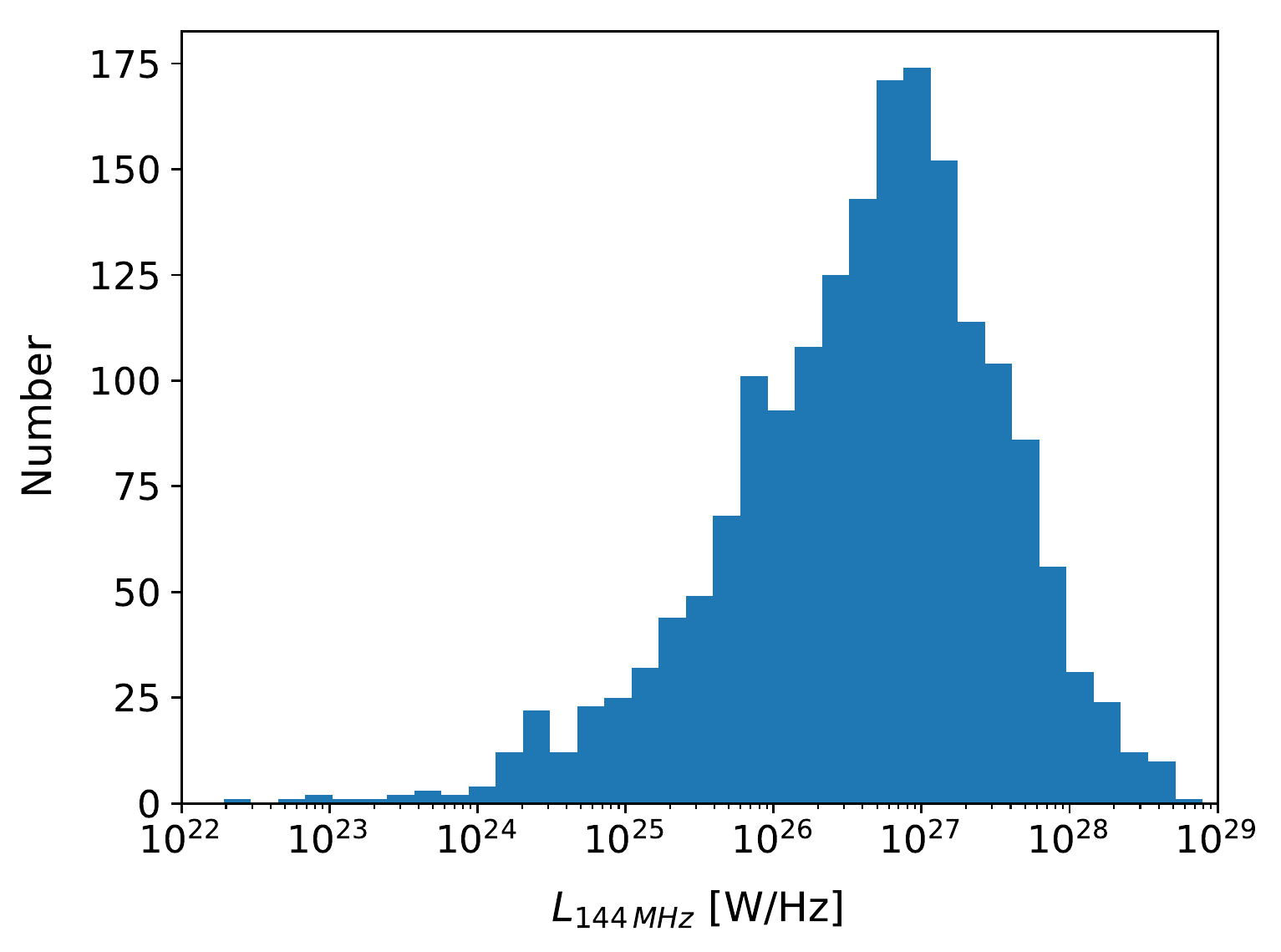}
\caption{ Histogram of the radio luminosity at 144~MHz, in W~Hz$^{-1}$, for all sources with a redshift estimate (either spectroscopic or photometric). The median 144~MHz luminosity is $\sim5\times10^{26}$~W~Hz$^{-1}$.}\label{fig:L150}
\end{figure}

For those sources with a redshift, the total flux from the LoTSS-DR2 catalogue and the largest angular size estimates are used to calculate the luminosity and linear size, assuming a flat $\Lambda$CDM cosmology with H$_0 = 67.8$ km s$^{-1}$ Mpc$^{-1}$ and $\Omega_M=0.308$ \citep{planck2016xiii}.
The spectral luminosity at 144 MHz ($L_{144\,{\rm MHz}}$) is estimated using a spectral index of $-0.7$ for all sources. The $L_{144\,{\rm MHz}}$ distribution is shown in Fig.~\ref{fig:L150} with a median spectral luminosity of $\sim5.3\times10^{26}$~W~Hz$^{-1}$. Two-thirds of the sample have $L_{144\,{\rm MHz}}$ above the traditional FRI/FRII luminosity boundary of $\sim10^{26}$~W~Hz$^{-1}$. 
 
\begin{figure}
\includegraphics[width=8.5cm,clip=true,trim=0.0cm 0.0cm 0.0cm 0.0cm]{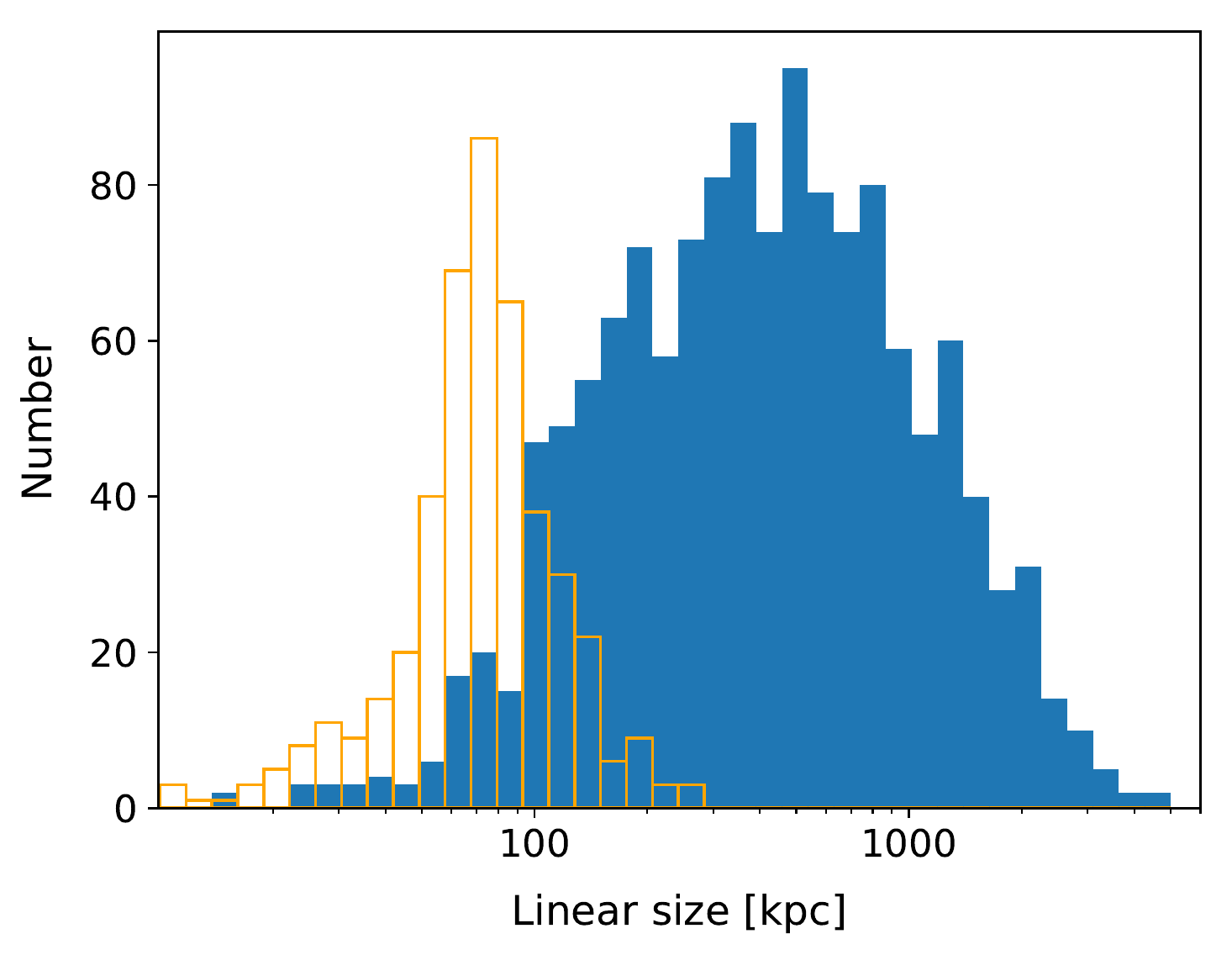}
\caption{ Histogram of the linear size, in kpc, of resolved sources (blue solid) and upper limits for compact sources (orange open), using spectroscopic or photometric redshifts. The median linear size is $\sim$300 kpc, while 266 sources have a linear size greater than 1 Mpc ($\sim$11\% of the sample). }\label{fig:ls}
\end{figure}

\begin{figure}
\includegraphics[width=8.5cm,clip=true,trim=0.0cm 0.0cm 0.0cm 0.0cm]{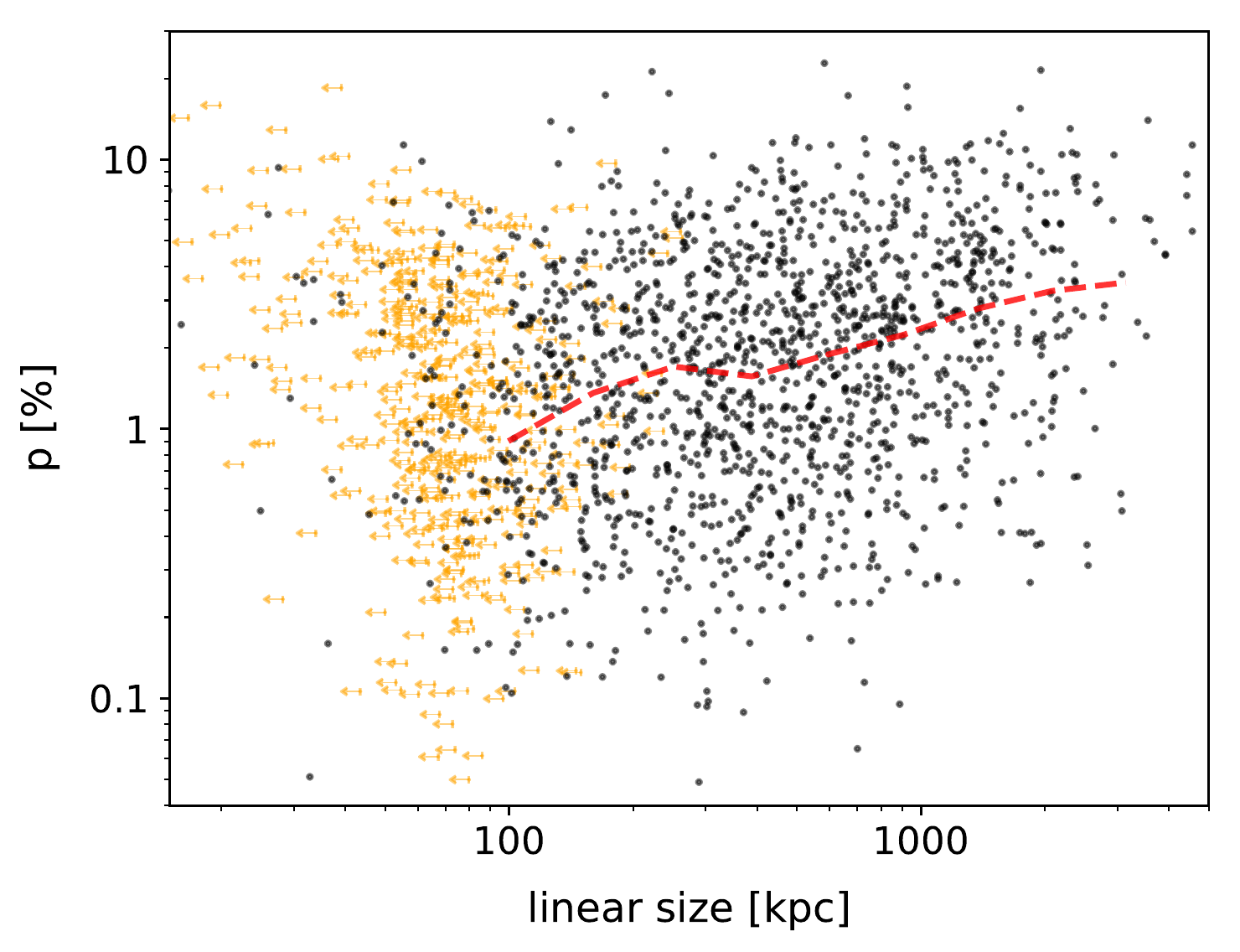}
\caption{The degree of polarization versus the projected linear size, with upper limits shown by orange arrows, and the median trend shown by the red dashed line (starting from where the upper limits are not the majority of points). }\label{fig:ls_p}
\end{figure}

The projected linear size distribution for the resolved sources is shown in Fig.~\ref{fig:ls}, with the median linear size being $\sim$400~kpc. 
The largest angular sizes of the sources are taken from either the MKSP-LGZ effort or from the major axis of the source as defined by the source finder {\sc pybdsf}. We only use sources that are resolved ($\sim$73\% of the RM Grid catalogue), defined by the same criteria as in \cite{shimwell2022}. Including upper limits for the unresolved sources leads to a median linear size for the RM Grid sources of $\sim$300~kpc, highlighting that the majority of polarized sources at low frequencies are large radio galaxies. 
There are 284 (457) sources that have an estimated linear size greater than 1 Mpc (0.7 Mpc). The three largest sources in the sample 
are 3C\,236 at 
4.38~Mpc, ILT~J093121.58+320211.0 at 4.2~Mpc, and ILT~J123459.82+531851.0 at 3.4~Mpc \citep{osullivan2019}. 
The faintest, and one of the smallest radio galaxies in the sample, is NGC\,5322, with $L_{\rm 144 MHz} \sim 2.3\times10^{22}$~W~Hz$^{-1}$ and a linear size of $\sim$26~kpc, falling into the category of a ``galaxy-scale jet'' \citep{webster2021}. 
The detection in polarization of a galaxy-scale jet is somewhat surprising as the influence of Faraday depolarization is expected to be quite strong on small scales within host galaxy halos \citep[e.g.][]{stromjaegers1988}, and because to date, most polarized radio galaxies found at low frequencies have large linear sizes \citep{osullivan2018b}. However, the avoidance of galaxy cluster environments and their relatively large angular size due to their location in the local Universe can provide more favourable conditions for the detection of polarized emission at low frequencies. In general, the degree of polarization increases with linear size (Fig.~\ref{fig:ls_p}), with a median degree of polarization of 0.9\% at 100~kpc and 2.4\% at 1~Mpc. 

The estimated spectral luminosity and linear sizes can be improved using {\sc lomorph}\footnote{https://github.com/bmingo/LoMorph}, as described in \cite{mingo2019}, by obtaining more robust estimates of the largest angular size and the total flux of the sources. It is known that the basic catalogue approaches used here can overestimate FRII sizes and potentially underestimate FRI sizes, as well as underestimating the total flux in general \citep{mingo2019}. The radio morphology class can also be obtained using {\sc lomorph}, and a preliminary analysis indicates that the largest class of polarized source are FRII ($\sim$40\%).  
The rest of the sources are classified as FRI ($\sim$20\%), hybrid ($\sim$15\%) and a combination of compact and unresolved ($\sim$25\%). 
It is striking to note the difference in the number of FRII relative to FRI ($\sim$2x) for polarized sources compared to estimates for the general population (i.e.~bright, large radio AGN), where the opposite is the case, with 2 to 3 times more FRIs than FRIIs \citep{mingo2019}. 

\subsubsection{Blazars}\label{sec:blazars}
Cross-matching the RM Grid catalogue with the ROMA-BZCAT all-sky blazar catalogue \citep{massaro2015} enabled us to identify 172 known blazars ($\sim$7\% of the RM Grid catalogue), for which 150 have redshifts. There are 64 BL Lacs and 100 FSRQs, with 8 classified as having an `uncertain blazar type'. 
The detection fraction of blazars is similar to that in DR1 RM catalogue \citep{vaneck2018}, where $\sim$10\% of sources were blazars \citep{osullivan2018b}.

\subsection{The nature of polarized sources at MHz frequencies}
This work allows us to comment on some specific differences between the types of polarized sources found at MHz compared to GHz frequencies. 
One of the most striking aspects of polarized sources at 144 MHz is that the majority have FRII morphologies, and there are twice as many FRIIs as FRIs. 
In contrast, at 1.4 GHz there are similar numbers of FRI and FRII polarized sources \citep{osullivan2015, banfield2011}. There are at least a couple of explanations for this. 
Firstly, FRIs are more commonly found in galaxy clusters \citep{best2009,gendre2013,croston2019} and the strong Faraday depolarization observed there \citep[e.g.][]{osinga2022} means that polarized sources embedded in (and in the background of) these environments are not found at 144 MHz  \citep{carretti2022, carretti2022b}. 
Secondly, the brightest and most highly polarized regions of FRIIs are at the source extremities, extending well beyond the host galaxy environment in many cases, where they are known to experience lower depolarization \citep[e.g.][]{stromjaegers1988}. 
The compact nature of FRII hotspots also means that the variation in RM across the emission region will be relatively small, minimising the amount of depolarization. 
In general, these low-depolarization (or `Faraday-simple') sources typically found at MHz frequencies makes them excellent probes of the low gas density and weak magnetic field environments of cosmic web filaments and voids \citep[e.g.][]{stuardi2020,pomakov2022,carretti2022b}. 
In contrast to studies at GHz frequencies, we have not found any polarized emission from star-forming galaxies, only radio-loud AGN, presumably because of the strong depolarization experienced by synchrotron emission in disk galaxies \citep[e.g.][]{beck2015}. 

However, it is not exclusively large, well-resolved radio galaxies that are detected. 
There is a significant population of compact and unresolved polarized sources detected ($\sim$25\% of the sample), with linear sizes $\lesssim100$~kpc (Section~\ref{sec:morph}). 
Further investigation is required to determine the exact nature of these sources as only about a quarter of them are known blazars (Section~\ref{sec:blazars}). The integrated emission from blazars can often exhibit complex spectral behaviour in both total intensity and polarization, due to the contribution of multiple inner jet components with different spectral and polarization properties, and thus the derived RMs from these sources should be considered with caution \citep[e.g.][]{osullivan2012}. 

The remaining compact sources could be small FRI/II sources \citep{capetti2017a,capetti2017b}, ``FR0'' sources \citep{baldi2018}, and/or Peaked Spectrum (PS) sources \citep{odea2021}. For these compact sources, Faraday complexity as well as spectral index effects can be important, where multiple bright components with differing spectral and polarization properties can contribute to polarization angle rotations not exclusively due to Faraday rotation \citep[][]{burn1966}, with spectral variations particularly important for the PS sources \citep[e.g.][]{ross2022}. The smallest PS sources are typically strongly depolarized \citep{cotton2003}, however many show polarization that survives to long wavelengths, even repolarizing in some cases \citep{mantovani2009}. 
Detailed spectral studies of the LoTSS sources, in addition to 0.3" imaging to determine their morphology, will allow us to resolve many of these uncertainties. Such studies may also provide new insights on the physical nature and environment of these sources, by incorporating the polarization information into models of free-free absorption and synchrotron self-absorption effects \citep[e.g.][]{callingham2017}. 

\subsection{Pulsar identifications}\label{sec:pulsars}
Many pulsars have emission that is highly linearly polarized \citep[e.g.][]{gouldlyne1998}, such that candidate pulsars can sometimes be identified in radio continuum 
images as unresolved, steep-spectrum, highly polarized sources \citep[e.g.][]{navarro1995}. Finding new pulsars is important because, for example, pulsars are useful 
for understanding the Galactic population of neutron stars, as probes of fundamental physics, and as precision 
probes of the Milky Way magnetic field using the combination of the RM and the dispersion measure \citep[e.g.][]{sobey2019}. 
We found 25 pulsars in the DR2 area, 24 of which were previously known, with one being a new discovery, which is described in \cite{sobey2022}. 
The 25 pulsars are listed in Table~\ref{tab:pulsars}, providing new high precision RMs and positions. 

\section{Summary}\label{sec:summary}

We have produced an RM Grid catalogue from the LoTSS-DR2 data containing 2461 RM values over an area of 5720~deg$^{2}$ (Table~\ref{tab:RMall}). This is the largest low-frequency RM Grid to date. 

\begin{enumerate}

\item The catalogue is derived from two non-contiguous sky areas: the 13~hr field which has an area of 4240~deg$^{2}$ with a polarized source areal number density of 0.48~deg$^{-2}$, and the 0~hr field with an area of 1480~deg$^{2}$ and source density of 0.29~deg$^{-2}$. \\

\item The RM values were derived from the LoTSS $Q$ and $U$ images at an angular resolution of 20\arcsec~across a frequency range of 120 to 168~MHz. The Faraday depth range was limited to $\pm120$\rad~and only polarized sources above $8\sigma_{\rm QU}$ were included in the catalogue. The median value of $\sigma_{\rm QU}$ in the Faraday spectra of the detected sources across 844 individual LoTSS-DR2 pointings was 0.08~mJy. \\

\item The typical FWHM of the Rotation Measure Spread Function (RMSF) is $\sim$1.16\rad, and the median RM uncertainty is $\sim$0.06\rad. The residual RM (RRM), after subtraction of the Galactic RM model, has a standard deviation of $\sim$2\rad~and is not correlated with the degree of polarization. \\

\item Of the $\sim$1.2 million LoTSS-DR2 sources with a peak total intensity greater than 1~mJy~beam$^{-1}$, only $\sim$0.2\% were polarized above $8\sigma_{\rm QU}$, with a median polarized intensity of 1.7 mJy~beam$^{-1}$. The degree of polarization of the detected sources ranges from 0.05\% to 31\%, with a median value of 1.8\%. \\

\item Only 37\% of the LoTSS RM Grid catalogue have a corresponding RM value in the NVSS RM catalogue (because LoTSS is much more sensitive for steep spectrum sources), with 90\% of those RM values consistent within $3\sigma$. \\

\item Host galaxy identifications were found for 88\% of the sources, leading to redshift estimates for 79\% of the sample (both spectroscopic and photometric). The median redshift is 0.6. The RRM is flat as a function of redshift, while the degree of polarization decreases by a factor of $\sim$10 between $z=0$ and $z=3$.\\

\item The median linear size of all polarized sources is $\sim$300~kpc 
and the median luminosity at 144~MHz is $\sim5\times10^{26}$~W~Hz$^{-1}$. The median degree of polarization increases with linear size, from 0.9\% at 100~kpc to 2.4\% at 1~Mpc. \\

\item The dominant radio morphology class is FRII ($\sim$40\% of sources), with $\sim$20\% FRI, $\sim$15\% hybrid and $\sim$25\% compact and unresolved sources. There are 172 polarized sources identified with known blazars. \\

\item We identified 25 pulsars that appeared as highly linearly polarized sources in our data (Table~\ref{tab:pulsars}). 
These sources are excluded from the RM Grid catalogue of 2461 sources. \\

\end{enumerate}

\subsection{Future LoTSS RM Grid enhancements}\label{sec:future}
The LoTSS-DR2 results demonstrate the potential of the LoTSS survey to produce a high quality RM Grid. However, there are several improvements that can be made to enhance the quality and areal number density of the RM Grid as the LoTSS survey continues. The next advance will be the production of $Q$ and $U$ image cubes at 6\arcsec~resolution (an improvement by a factor of 3.3), which should increase the polarized source areal number density, primarily by reducing the effects of beam depolarization. 
For example, the majority of polarized sources detected at 20\arcsec~are resolved (Section~\ref{sec:morph}), hence resolving more sources will increase the chances of new polarized source detections, in addition to providing new polarized source components for already detected sources, which are valuable for e.g.~RM pair studies \citep{pomakov2022}. 
In addition to this, RM synthesis will be run without any masking (the data were masked at 1~mJy~beam$^{-1}$ in Stokes $I$ in this work), providing better sensitivity to highly polarized sources, which further enhances the discovery potential of the survey, and will enable a better quantification of the noise properties in polarization for each field. 

Extending the Faraday depth range to search for polarized sources with high $|$RM$|$ values (i.e.~$>120$\rad) will also be included, which becomes more important for fields which are closer to the Galactic plane. Ideally, a finer channelisation would be used for the Galactic plane region (i.e.~48~kHz for a max/min RM of $\pm$900\rad), to decrease the effects of bandwidth depolarization. In the longer term, a deconvolution strategy for $Q$ and $U$ needs to be implemented. 

In terms of the polarization calibration of the data, enhancements to reduce the widefield instrumental polarization without compromising the polarization data reliability are needed. An absolute polarization angle calibration strategy for each field, similar to that presented in \cite{herreraruiz2021}, would allow mosaicking of the data and thus enable us to obtain the maximum sensitivity of the LoTSS survey for polarization. Improvements in the ionosphere RM correction are also desirable since the RM errors are dominated by the residual errors in the ionosphere RM correction (i.e.~$\sim$5 to 10 times larger than the measurement errors). The continuing advances in using the full LOFAR array to achieve an angular resolution of 0.3\arcsec~\citep[e.g.][]{morabito2022,sweijen2022} has great potential for finding many more polarized sources to further enhance the areal number density of the LoTSS RM Grid. 

\section*{Acknowledgments}
SPO and MB acknowledge financial support from the Deutsche Forschungsgemeinschaft (DFG) under grant BR2026/23. 
MB acknowledges support from the Deutsche Forschungsgemeinschaft under Germany's Excellence Strategy - EXC 2121 ``Quantum Universe'' - 390833306. 
AMS gratefully acknowledges support from an Alan Turing Institute AI Fellowship EP/V030302/1. 
KJD acknowledges funding from the European Union's Horizon 2020 research and innovation programme under the Marie Sk\l{}odowska-Curie grant agreement No. 892117 (HIZRAD). 
MJH acknowledges support from the UK STFC [ST/V000624/1]. 
JHC and BM acknowledge support from the UK Science and Technology Facilities Council (STFC) under grants ST/R000794/1, and ST/T000295/1
M.~Bilicki is supported by the Polish National Science Center through grants no.~2020/38/E/ST9/00395, 2018/30/E/ST9/00698, 2018/31/G/ST9/03388 and 2020/39/B/ST9/03494, and by the Polish Ministry of Science and Higher Education through grant DIR/WK/2018/12.
VV acknowledges support from INAF mainstream project ``Galaxy Clusters Science with LOFAR'' 1.05.01.86.05.
BNW acknowledges support from the Polish National Science Centre (NCN), grant no.~UMO-2016/23/D/ST9/00386. The authors of the Polish scientific institutions thank the Ministry of Science and Higher Education (MSHE), Poland for granting funds for the Polish contribution to the ILT (MSHE decision no.~DIR/WK/2016/2017/05-1) and for maintenance of the LOFAR LOFAR PL-611 Lazy, LOFAR PL-612 Baldy stations (MSHE decisions: no.~46/E-383/SPUB/SP/2019 and no.~59/E-383/SPUB/SP/2019.1, respectively). 
AD acknowledges support by the BMBF Verbundforschung under the grant 05A20STA. 
RJvW acknowledges support from the VIDI research programme with project number 639.042.729, which is financed by the Netherlands Organisation for Scientific Research (NWO). 
This research made use of Astropy, a community-developed core Python package for astronomy \citep{astropy2013} hosted at http://www.astropy.org/, of Matplotlib \citep{hunter2007}, of APLpy \citep{aplpy2012}, an open-source astronomical plotting package for Python hosted at http://aplpy.github.com/, and of TOPCAT, an interactive graphical viewer and editor for tabular data \citep{taylor2005}. This research has made use of ``Aladin sky atlas'' developed at CDS, Strasbourg Observatory, France \citep{aladin}. 
LOFAR \citep{vanhaarlem2013} is the Low Frequency Array designed and constructed by ASTRON. It has observing, data processing, and data storage facilities in several countries, which are owned by various parties (each with their own funding sources), and that are collectively operated by the ILT foundation under a joint scientific policy. The ILT resources have benefited from the following recent major funding sources: CNRS-INSU, Observatoire de Paris and Universit\'e d'Orl\'eans, France; BMBF, MIWF-NRW, MPG, Germany; Science Foundation Ireland (SFI), Department of Business, Enterprise and Innovation (DBEI), Ireland; NWO, The Netherlands; The Science and Technology Facilities Council, UK; Ministry of Science and Higher Education, Poland; The Istituto Nazionale di Astrofisica (INAF), Italy.
This research made use of the Dutch national e-infrastructure with support of the SURF Cooperative (e-infra 180169) and the LOFAR e-infra group. The J\"ulich LOFAR Long Term Archive and the German LOFAR network are both coordinated and operated by J\"ulich Supercomputing Centre (JSC), and computing resources on the supercomputer JUWELS at JSC were provided by the Gauss Centre for Supercomputing e.V.~(grant CHTB00) through the John von Neumann Institute for Computing (NIC).
This research made use of the University of Hertfordshire high-performance computing facility and the LOFAR-UK computing facility located at the University of Hertfordshire and supported by STFC [ST/P000096/1], and of the Italian LOFAR IT computing infrastructure supported and operated by INAF, and by the Physics Department of Turin university (under an agreement with Consorzio Interuniversitario per la Fisica Spaziale) at the C3S Supercomputing Centre, Italy. The authors thank the referee for comments which improved the paper. 

\section*{Data Availability}\label{datalinks}
The RM Grid catalogue (and the $Q$ and $U$ frequency spectra for each catalogue entry) can be downloaded from the public MKSP website at \url{https://lofar-mksp.org/data/}. The catalogues are also available through the Virtual Observatory at \url{https://dc.g-vo.org/}. 
The LoTSS $QU$ frequency cubes at 20" and the RM synthesis output are stored on the SURF Data Repository at \url{https://repository.surfsara.nl/} with DOI:10.25606/SURF.lotss-dr2. 

\bibliographystyle{mnras.bst}
\bibliography{igmf} % if your bibtex file is called igmf.bib

\appendix

\vspace{-0.5cm}
\section{RM Grid column descriptions}\label{columns}
The RM catalogue generally follows the RMTable standard\footnote{https://github.com/CIRADA-Tools/RMTable}, as described in Van Eck et al.~in prep., with some additional value-added columns (e.g.~LoTSS-DR2 total intensity source associations, host galaxy coordinates, redshift, etc.). 
Each row in the catalogue corresponds to a single polarized component. \\

\noindent cat\_id: unique identifier for each polarized component. \\
ra: Right Ascension (J2000) of the polarized component in degrees.\\
dec: Declination (J2000) of the polarized component in degrees.\\
pos\_err: Positional uncertainty in degrees. \\
rm: Faraday rotation measure value at the peak of the Faraday dispersion function in\rad~(excluding sources that are due to instrumental polarization, see Section~\ref{sec:original}).\\
rm\_err: Uncertainty in the peak RM value in\rad, including some systematic errors.\\
rm\_err\_snr: Uncertainty in the peak RM value in\rad, based only on the signal-to-noise ratio.\\
polint: Linear polarization intensity at the reference frequency, after correction for polarization bias, in Jy~beam$^{-1}$. \\
polint\_err: Uncertainty in the linear polarization intensity in Jy~beam$^{-1}$. \\
fracpol: Fractional linear polarization of the polarized component. \\
fracpol\_err: Uncertainty in the fractional linear polarization. \\
stokesI: Total intensity at the position of the polarized component in Jy~beam$^{-1}$. \\
stokesI\_err: Uncertainty in the total intensity in Jy~beam$^{-1}$. \\
rmsf\_fwhm: Full width at half maximum of the rotation measure spread function (RMSF) in\rad.\\
reffreq\_pol: Reference frequency for the linear polarization quantities in Hz.\\
reffreq\_I: Reference frequency for the total intensity values in Hz.\\
Nchan: The number of Stokes $Q$ and $U$ image planes (i.e.~channels) used in RM synthesis.\\
noise\_chan: The median noise of the Stokes $Q$ and $U$ image planes in Jy~beam$^{-1}$. \\
epoch: MJD for the observation of the corresponding LoTSS field. \\
int\_time: Integration time for each field in seconds.\\
p\_orig: original polarized intensity from selection of candidate sources (as described in Section~\ref{sec:original}) in Jy~beam$^{-1}$. \\
snr\_orig: signal to noise ratio in polarization from original selection of candidate sources (c.f.~Section~\ref{sec:original}). \\
source\_name\_DR2: ILT name of the LoTSS-DR2 total intensity source associated with the polarized component.\\
RA\_DR2: Right Ascension (J2000) of the total intensity source in degrees.\\
E\_RA\_DR2: Uncertainty in the Right Ascension in degrees. \\
DEC\_DR2: Declination (J2000) of the total intensity source in degrees.\\
E\_DEC\_DR2: Uncertainty in the Declination in degrees. \\
Total\_flux\_DR2: Integrated flux density of the LoTSS-DR2 source in Jy.\\
E\_Total\_flux\_DR2: Uncertainty in the integrated flux density in Jy.\\
Maj\_DR2: Major axis size of the source in arcseconds.\footnote{The size of the sources with multiple Gaussian components are estimated by {\sc pybdsf} through a moment analysis \url{https://www.astron.nl/citt/pybdsf/algorithms.html\#grouping.}} \\
Min\_DR2: Minor axis size of the source in arcseconds. \\
PA\_DR2: Position angle of the source in degrees. \\
field: LoTSS field name. \\
ra\_centre: Right Ascension (J2000) of the centre of the field in degrees.\\
dec\_centre: Declination (J2000) of the centre of the field in degrees.\\
beamdist: Distance of the polarized component from the centre of the field in degrees.\\
x: x-pixel coordinate within an individual LoTSS field, ranging from 0 to 3200, for a pixel width of 4.5\arcsec. \\
y: y-pixel coordinate within an individual LoTSS field, ranging from 0 to 3200, for a pixel height of 4.5\arcsec. \\
l: Galactic Longitude of the polarized component in degrees. \\
b: Galactic Latitude of the polarized component in degrees. \\
LGZ\_Size: Largest angular size estimated from visual inspection. \\
lgz\_ra\_deg: Right Ascension (J2000) of the host galaxy in degrees.\\
lgz\_dec\_deg: Declination (J2000) of the host galaxy in degrees.\\
z\_best: redshift of host galaxy, spectroscopic if available, otherwise photometric. \\
phot\_spec\_z\_best: a value of 0/1 corresponds to a photometric/spectroscopic redshift in the ``z\_best'' column. \\
L144: Estimate of the luminosity at 144 MHz in W~Hz$^{-1}$.\\
linearsize\_kpc: Estimate of the projected largest linear size in kpc.\\
bzcat\_name: Blazar source name in the ROMA-BZCAT catalogue \citep{massaro2015}.\\
RRM2022\_1deg: The residual RM after subtraction of the average GRM with a disc of diameter 1 degree, using v2 of the Galactic Faraday rotation sky at \url{https://wwwmpa.mpa-garching.mpg.de/~ensslin/research/data/faraday2020.html}. 
GRM2022\_1deg: The average GRM with a disc of diameter 1 degree. 
GRMerr2022\_1deg: The uncertainty in the average GRM value. 

\begin{figure*}
\includegraphics[width=8.5cm,clip=true,trim=0.0cm 0.0cm 0.0cm 0.0cm]{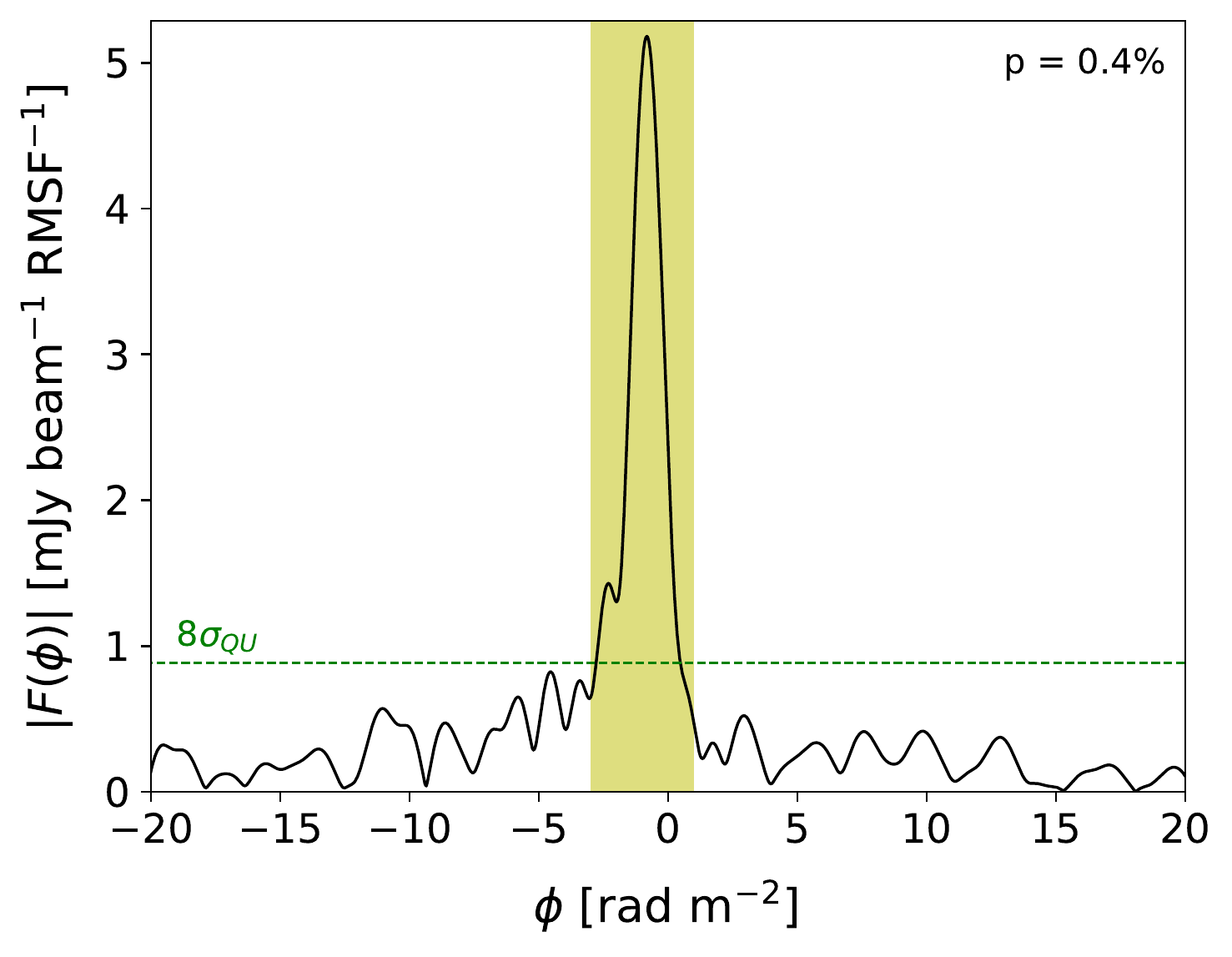}
\includegraphics[width=8.5cm,clip=true,trim=0.0cm 0.0cm 0.0cm 0.0cm]{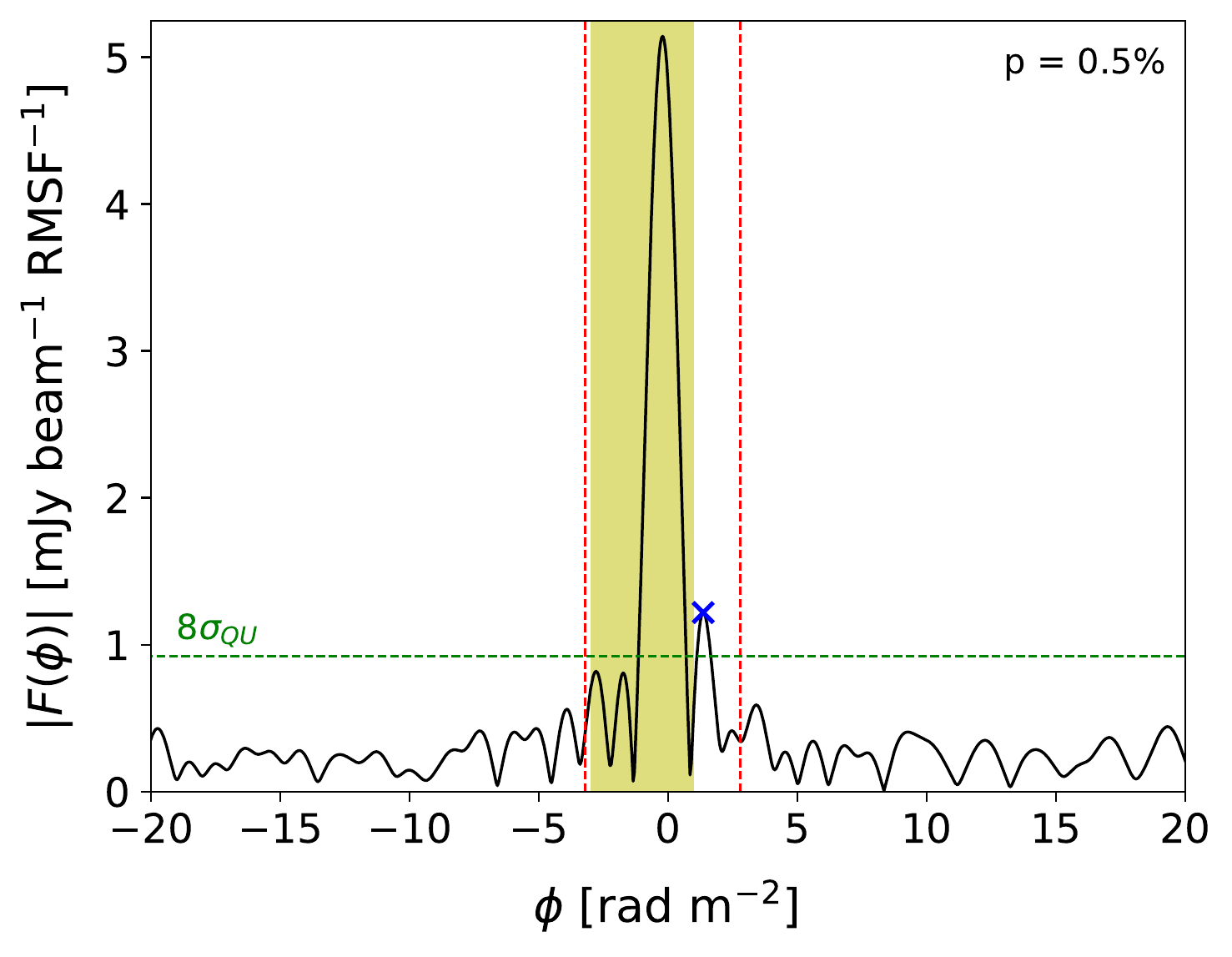}
\includegraphics[width=8.5cm,clip=true,trim=0.0cm 0.0cm 0.0cm 0.0cm]{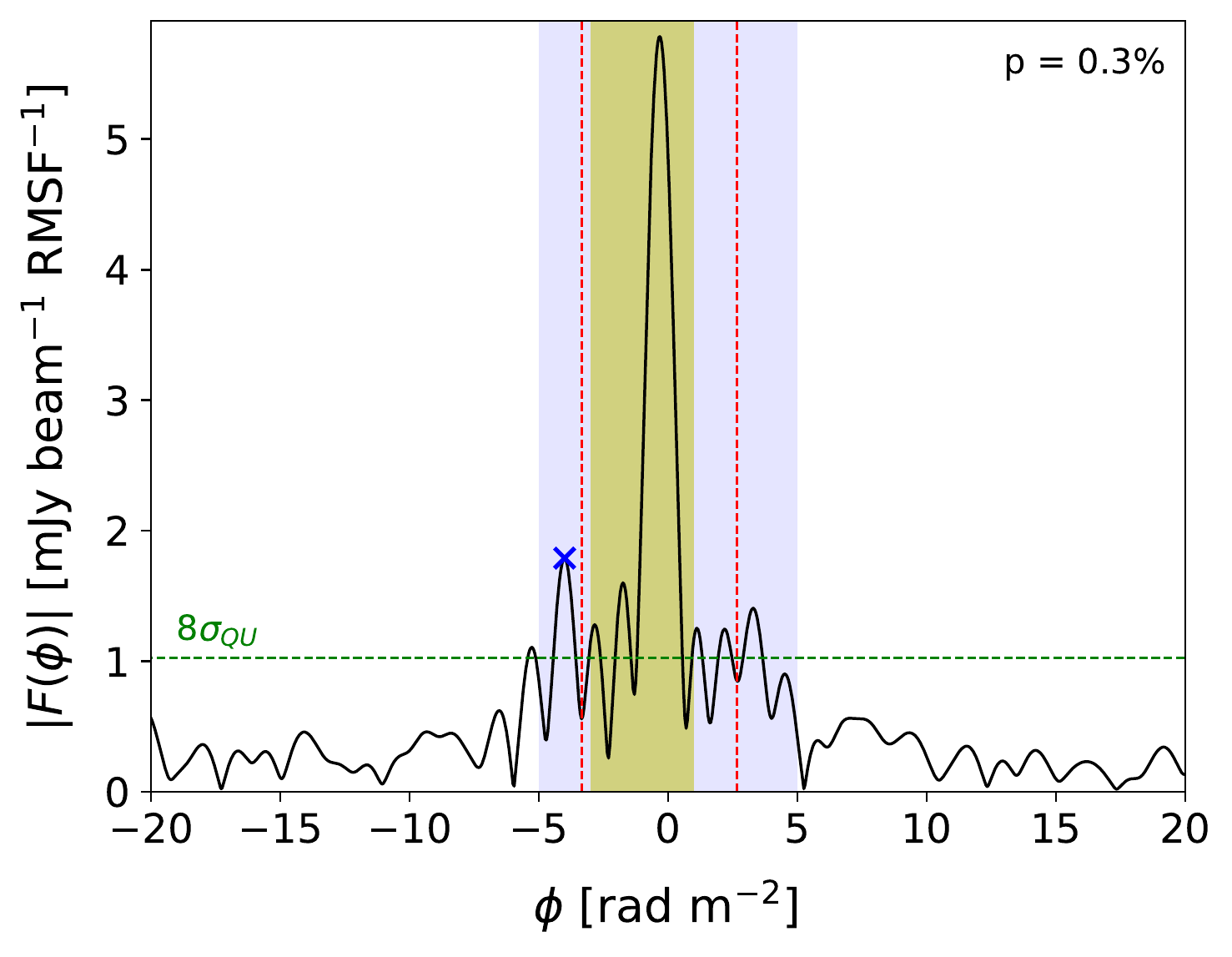}
\includegraphics[width=8.5cm,clip=true,trim=0.0cm 0.0cm 0.0cm 0.0cm]{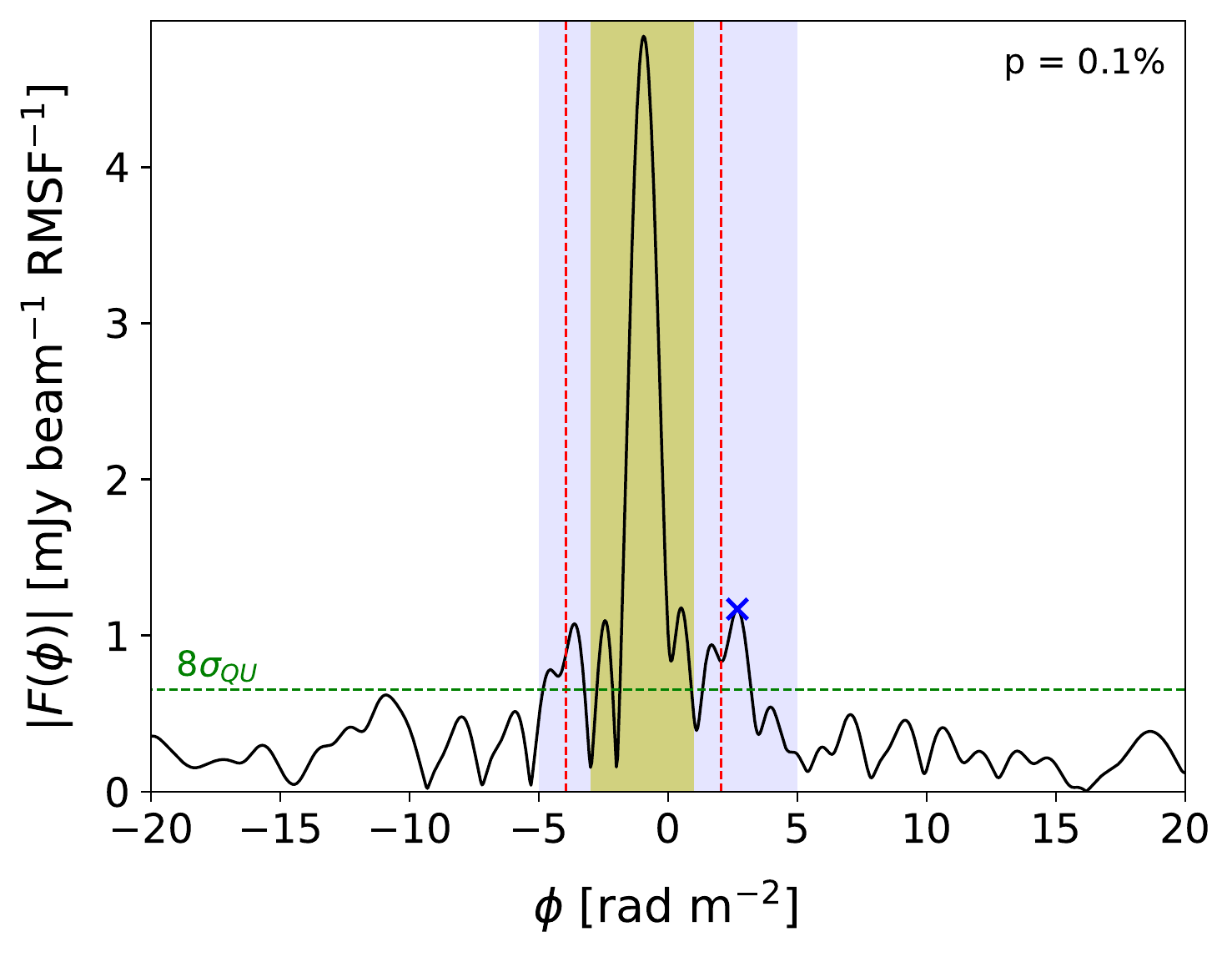}
\includegraphics[width=8.5cm,clip=true,trim=0.0cm 0.0cm 0.0cm 0.0cm]{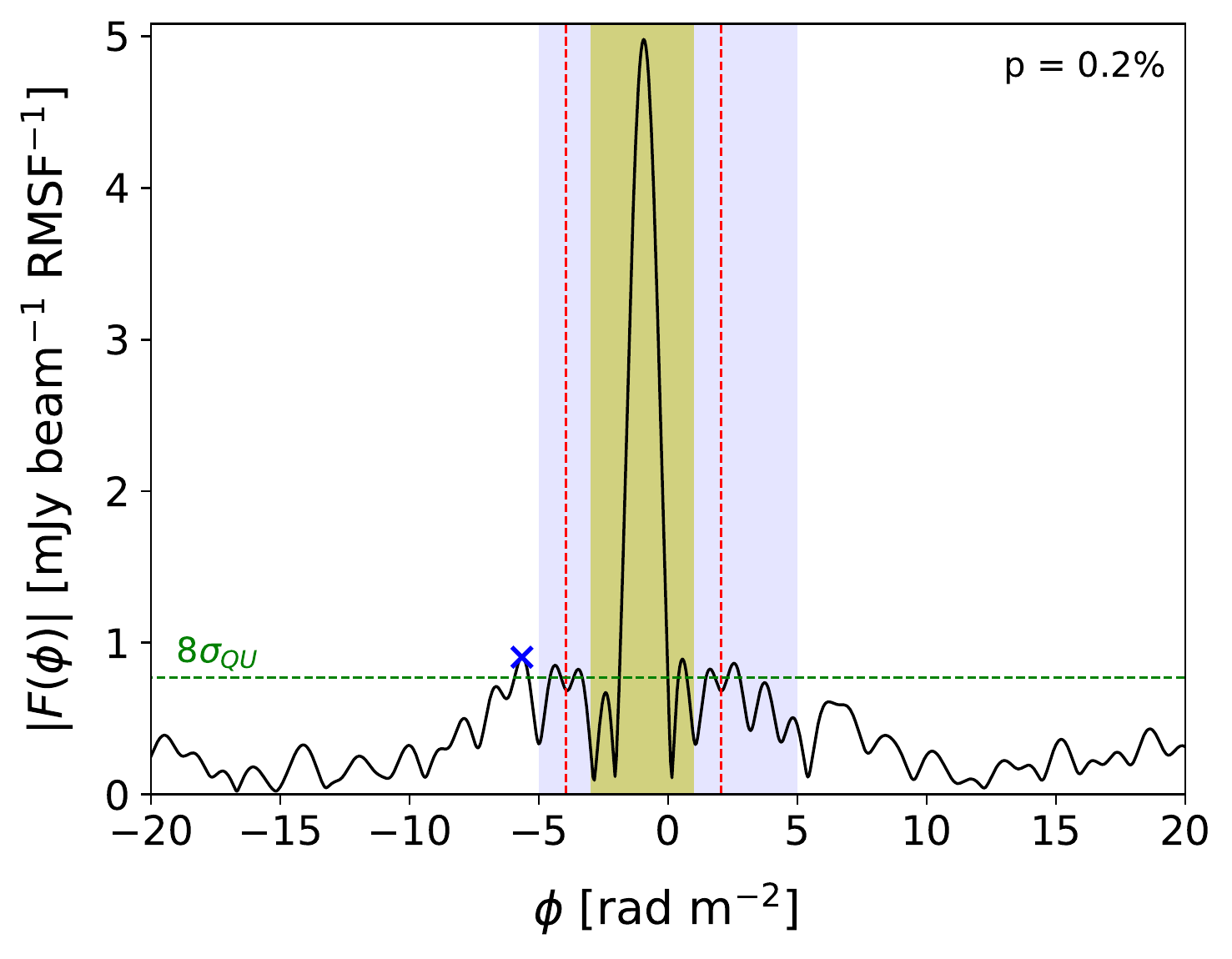}
\includegraphics[width=8.5cm,clip=true,trim=0.0cm 0.0cm 0.0cm 0.0cm]{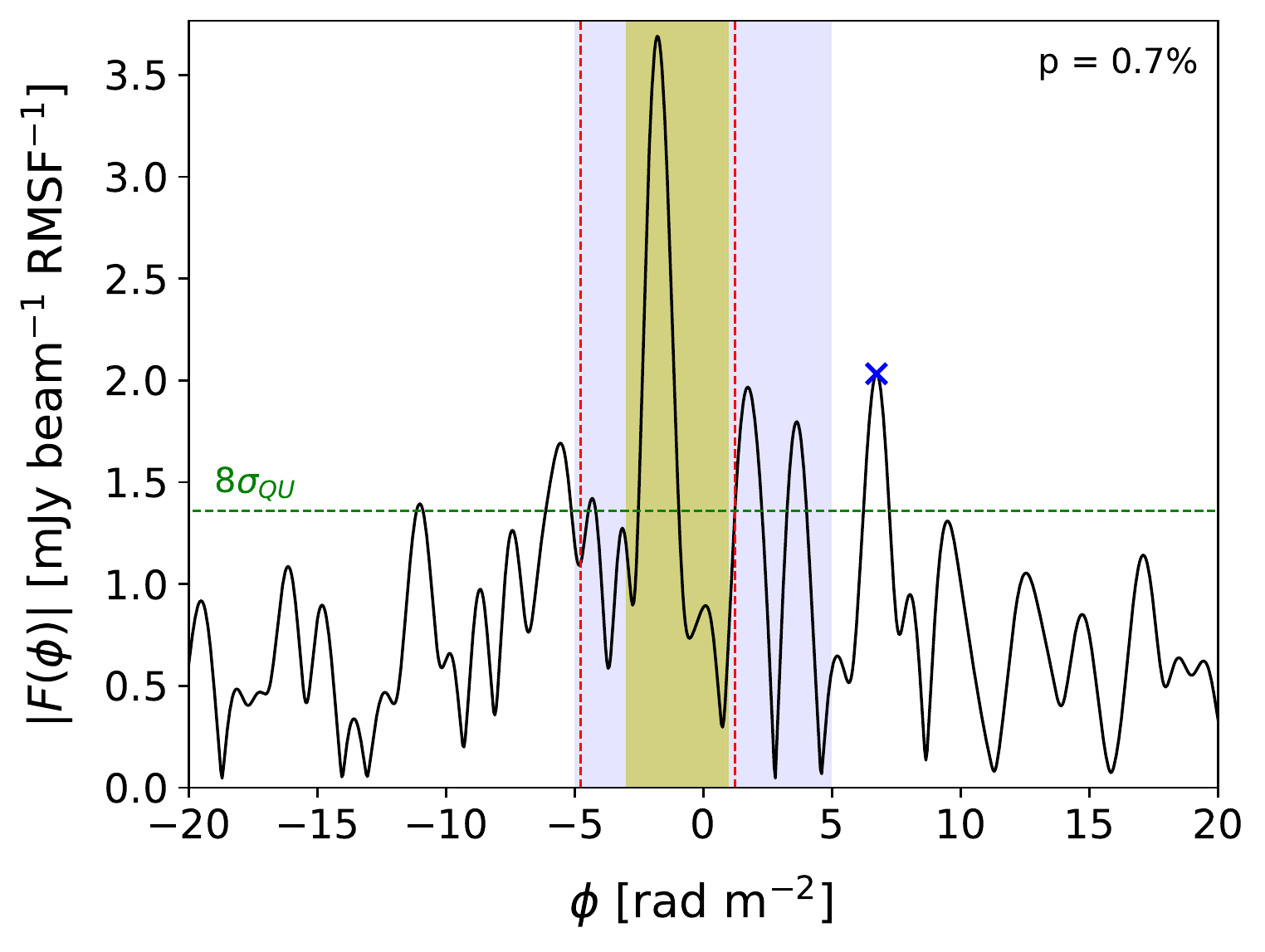}
\caption{Examples of sources excluded by the various leakage exclusion criteria employed on the FDF (as described in Section~\ref{sec:data}). The blue cross indicates the candidate polarized source that avoided the original leakage exclusion range. 
Top left: the original leakage exclusion range of $-3$\rad~to $+1$\rad, highlighted by the yellow shaded region. 
Top right: the extended leakage range of $\pm3$\rad~of the main leakage peak Faraday depth value, shown by vertical red dashed lines. 
Middle: two examples of an extended leakage exclusion range of $\pm5$\rad~for low degree of polarization sources, highlighted by the shaded blue region. 
Bottom: two examples of sources that escaped the above leakage exclusion criteria, but were excluded after visual inspection. 
The detection threshold of $8\sigma_{QU}$ is shown by the horizontal green dashed line and the degree of polarization is quoted in the top right corner of each plot. }\label{fig:leakage}
\end{figure*}

\section{Pulsars}\label{pulsars}
The list of the 25 pulsars, identified in the LoTSS-DR2 area through their high degree of linear polarization and unresolved total intensity source structure, are shown in Table~\ref{tab:pulsars}. 

\begin{table*}
\centering
\caption{List of pulsars identified in the LoTSS-DR2 sky area. }
\begin{tabular}{lcccccccc}
\hline
PSR name & RA      & Dec              & RM   &   $|\bm{F}(\phi)|$        &  $p$     &    Source Name &  LoTSS  & cat\_id \\
               & [J2000]  &  [J2000]        & [\rad] &  [mJy~beam$^{-1}$]  &   [\%]  &     LoTSS-DR2     &   Field       &    \\
\hline
J0154+1833 & 01:54:36.9 & 18:33:51 & -22.845 $\pm$ 0.056 & 2.45 $\pm$ 0.11 & 62.8 $\pm$ 2.8 & ILTJ015436.89+183350.8 & P029+19 & 4126 \\
J0158+21 & 01:58:46.0 & 21:06:47 & -29.771 $\pm$ 0.053 & 2.90 $\pm$ 0.09 & 65.5 $\pm$ 2.1 & ILTJ015845.99+210646.7 & P029+21 & 4130 \\
J0220+36 & 02:20:42.1 & 36:26:56 & -41.051 $\pm$ 0.052 & 4.56 $\pm$ 0.10 & 19.4 $\pm$ 0.4 & ILTJ022042.12+362655.7 & P035+36 & 4250 \\
B0751+32 & 07:54:40.6 & 32:31:56 & 4.732 $\pm$ 0.054 & 2.14 $\pm$ 0.08 & 16.9 $\pm$ 0.6 & ILTJ075440.65+323156.5 & P118+32 & 4796 \\
B0917+63 & 09:21:14.2 & 62:54:14 & -14.564 $\pm$ 0.058 & 1.18 $\pm$ 0.06 & 11.9 $\pm$ 0.6 & ILTJ092114.15+625413.9 & P141+62 & 5851 \\
J0944+4106 & 09:44:22.6 & 41:26:43 & 4.528 $\pm$ 0.063 & 0.86 $\pm$ 0.06 & 3.7 $\pm$ 0.2 & ILTJ094422.62+412642.5 & P146+42 & 214 \\
J1049+5822 & 10:49:37.9 & 58:22:18 & 3.950 $\pm$ 0.055 & 1.32 $\pm$ 0.05 & 63.5 $\pm$ 2.4 & ILTJ104937.86+582217.6 & P161+60 & 461 \\
B1112+50 & 11:15:38.5 & 50:30:11 & 2.633 $\pm$ 0.051 & 4.18 $\pm$ 0.07 & 5.6 $\pm$ 0.1 & ILTJ111538.50+503023.8 & P6 & 2117 \\
J1239+32 & 12:39:27.3 & 32:39:23 & -9.083 $\pm$ 0.051 & 3.19 $\pm$ 0.07 & 17.2 $\pm$ 0.4 & ILTJ123927.33+323923.4 & P188+32 & 708 \\
J1344+66 & 13:43:59.3 & 66:34:25 & 57.022 $\pm$ 0.063 & 0.73 $\pm$ 0.05 & 11.1 $\pm$ 0.7 & ILTJ134359.33+663425.3 & P205+67 & 7456 \\
B1508+55 & 15:09:25.1 & 55:31:33 & 1.430 $\pm$ 0.050 & 13.85 $\pm$ 0.16 & 2.0 $\pm$ 0.0 & ILTJ150925.49+553131.7 & P227+55 & 3004 \\
J1518+4904 & 15:18:16.8 & 49:04:34 & -12.306 $\pm$ 0.052 & 4.08 $\pm$ 0.10 & 69.1 $\pm$ 1.6 & ILTJ151816.80+490434.1 & P229+48 & 7999 \\
J1544+4937 & 15:44:04.5 & 49:37:55 & 9.916 $\pm$ 0.052 & 2.17 $\pm$ 0.06 & 34.5 $\pm$ 0.9 & ILTJ154404.49+493755.3 & P235+50 & 8145 \\
J1628+4406 & 16:28:50.2 & 44:06:43 & 3.361 $\pm$ 0.053 & 2.82 $\pm$ 0.09 & 27.5 $\pm$ 0.9 & ILTJ162850.24+440642.5 & P247+43 & 8360 \\
J1630+3550 & 16:30:35.9 & 35:50:43 & 8.522 $\pm$ 0.059 & 2.03 $\pm$ 0.11 & 32.6 $\pm$ 1.7 & ILTJ163035.93+355042.5 & P249+38 & 3246 \\
J1630+3734 & 16:30:36.5 & 37:34:42 & 1.726 $\pm$ 0.064 & 1.10 $\pm$ 0.08 & 35.2 $\pm$ 2.5 & ILTJ163036.45+373441.9 & P249+38 & 3245 \\
J1647+6608 & 16:47:32.5 & 66:08:22 & 7.758 $\pm$ 0.065 & 1.13 $\pm$ 0.08 & 18.8 $\pm$ 1.4 & ILTJ164732.46+660821.9 & P254+65 & 3277 \\
J1658+36 & 16:58:26.5 & 36:30:30 & 6.350 $\pm$ 0.050 & 13.52 $\pm$ 0.15 & 54.8 $\pm$ 0.6 & ILTJ165826.55+363030.4 & P255+38 & 3292 \\
J1722+35 & 17:22:09.5 & 35:19:19 & 33.111 $\pm$ 0.057 & 2.08 $\pm$ 0.10 & 17.9 $\pm$ 0.8 & ILTJ172209.51+351918.7 & P261+35 & 8662 \\
B1811+40 & 18:13:13.0 & 40:13:40 & 49.603 $\pm$ 0.053 & 5.74 $\pm$ 0.17 & 20.3 $\pm$ 0.6 & ILTJ181313.20+401339.3 & P272+40 & 8775 \\
J2212+24 & 22:12:27.6 & 24:50:37 & -35.314 $\pm$ 0.056 & 2.40 $\pm$ 0.11 & 31.2 $\pm$ 1.4 & ILTJ221227.63+245036.7 & P333+26 & 8993 \\
J2214+3000 & 22:14:38.8 & 30:00:38 & -44.531 $\pm$ 0.055 & 2.97 $\pm$ 0.12 & 57.1 $\pm$ 2.3 & ILTJ221438.84+300038.2 & P333+28 & 8999 \\
J2229+2643 & 22:29:50.9 & 26:43:57 & -57.259 $\pm$ 0.057 & 2.68 $\pm$ 0.12 & 26.4 $\pm$ 1.2 & ILTJ222950.89+264357.4 & P338+26 & 3466 \\
J2306+31 & 23:06:19.2 & 31:24:20 & -81.753 $\pm$ 0.051 & 4.50 $\pm$ 0.09 & 50.8 $\pm$ 1.0 & ILTJ230619.22+312420.2 & P348+31 & 3490 \\
B2315+21 & 23:17:57.9 & 21:49:47 & -37.698 $\pm$ 0.051 & 7.21 $\pm$ 0.11 & 18.1 $\pm$ 0.3 & ILTJ231757.85+214947.1 & P348+21 & 9226 \\
\hline
\end{tabular}
\label{tab:pulsars}
\end{table*}

\section*{Affiliations}
\noindent
{\it
$^{1}$ School of Physical Sciences and Centre for Astrophysics \& Relativity, Dublin City University, Glasnevin, D09 W6Y4, Ireland\\
$^{2}$ ASTRON, Netherlands Institute for Radio Astronomy, Oude Hoogeveensedijk 4, 7991 PD, Dwingeloo, The Netherlands\\
$^{3}$ Leiden Observatory, Leiden University, PO Box 9513, NL-2300 RA Leiden, The Netherlands\\
$^{4}$ Centre for Astrophysics Research, Department of Physics, Astronomy and Mathematics, University of Hertfordshire, College Lane, Hatfield AL10 9AB, UK\\
$^{5}$ GEPI, Observatoire de Paris, Universit\'e PSL, CNRS, 5 Place Jules Janssen, 92190 Meudon, France\\
$^{6}$ Department of Physics \& Electronics, Rhodes University, PO Box 94, Grahamstown, 6140, South Africa\\
$^{7}$ CSIRO Space and Astronomy, PO Box 1130, Bentley, WA 6102, Australia\\
$^{8}$ INAF, Istituto di Radioastronomia, Via Gobetti 101, 40129 Bologna, Italy\\
$^{9}$ University of Hamburg, Hamburger Sternwarte, Gojenbergsweg 112, D-21029, Hamburg, Germany\\
$^{10}$ INAF-Osservatorio Astronomico di Cagliari, Via della Scienza 5, I-09047 Selargius (CA), Italy\\
$^{11}$ Dunlap Institute for Astronomy and Astrophysics, University of Toronto, 50 St. George Street, Toronto, ON M5S 3H4, Canada\\
$^{12}$ Department of Space, Earth and Environment, Chalmers University of Technology, Onsala Space Observatory, 439 92 Onsala, Sweden\\
$^{13}$ Max-Planck-Institut f\"ur Radioastronomie, Auf dem H\"ugel 69, 53121 Bonn, Germany\\
$^{14}$ Center for Theoretical Physics, Polish Academy of Sciences, al. Lotnik\'ow 32/46, 02-668 Warsaw, Poland\\
$^{15}$ Overstock Ireland Ltd, Westgate, Finisklin Business Park, Sligo, Ireland F91 HF66\\
$^{16}$ School of Physical Sciences, The Open University, Walton Hall, Milton Keynes, MK7 6AA, UK\\
$^{17}$ Th\"uringer Landessternwarte, Sternwarte 5, D-07778 Tautenburg, Germany\\
$^{18}$ Institute for Astronomy, Royal Observatory, Blackford Hill, Edinburgh, EH9 3HJ, UK\\
$^{19}$ National Astronomical Observatory of Japan, 2-21-1 Osawa, Mitaka, Tokyo 181-8588, Japan\\
$^{20}$ Astronomical Observatory, Jagiellonian University, ul.~Orla 171, PL 30-244 Krak\'ow, Poland\\
$^{21}$ Jodrell Bank Centre for Astrophysics, Department of Physics \& Astronomy, University of Manchester, Oxford Road, Manchester M13 9PL, UK\\
$^{22}$ The Alan Turing Institute, Euston Road, London NW1 2DB, UK\\

\bsp
\label{lastpage}
\end{document}